%% file: main.tex
\documentclass[11pt]{article}

\input{setup/packages.tex}

\begin{document}

\thispagestyle{empty}
\setlength\headheight{0pt} 
\newgeometry{bottom=2cm, top=2cm} 
\begin{center}

\begin{center}
\includegraphics[width=0.4\linewidth]{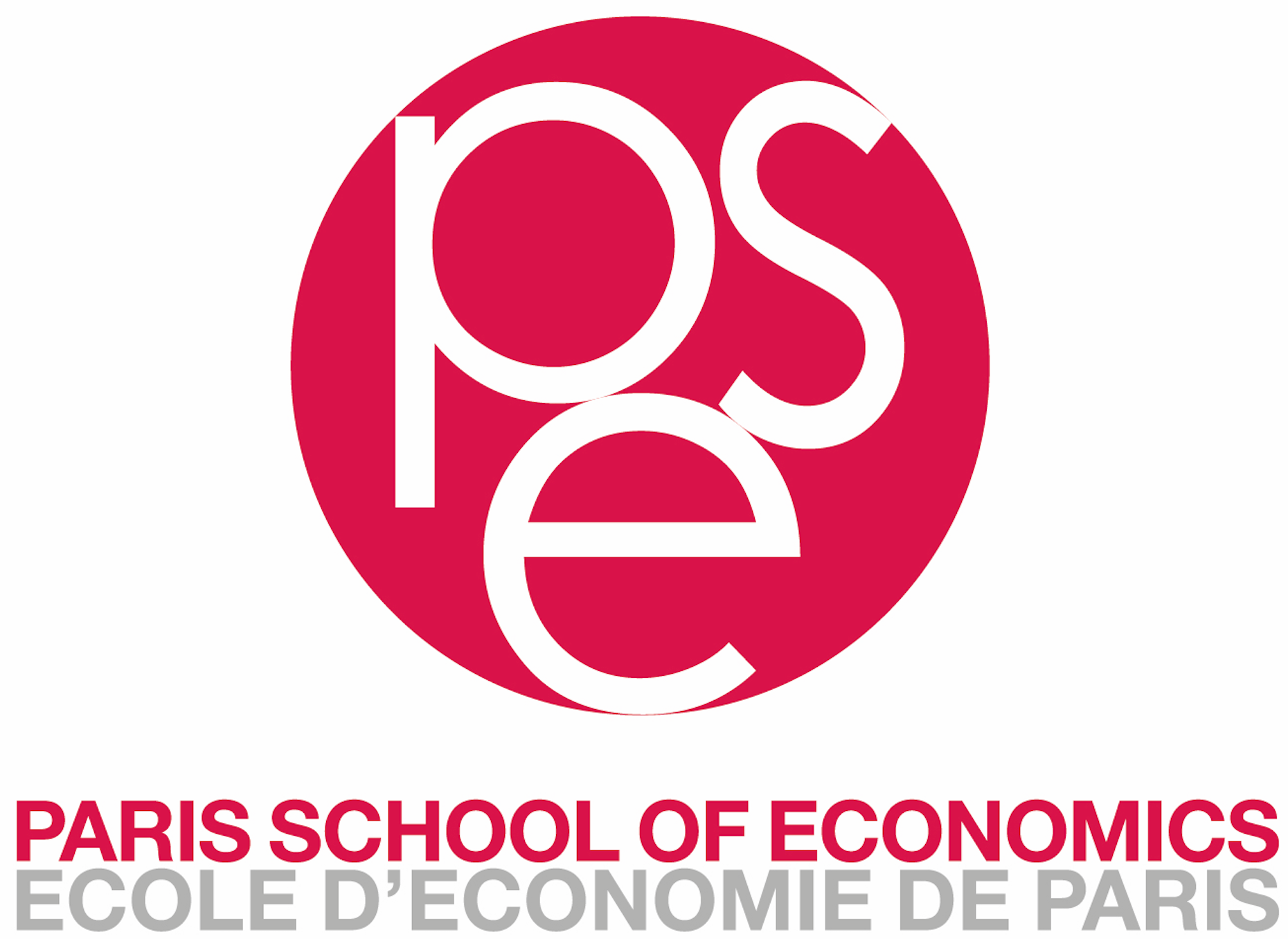}\;   
\includegraphics[width=0.28\linewidth]{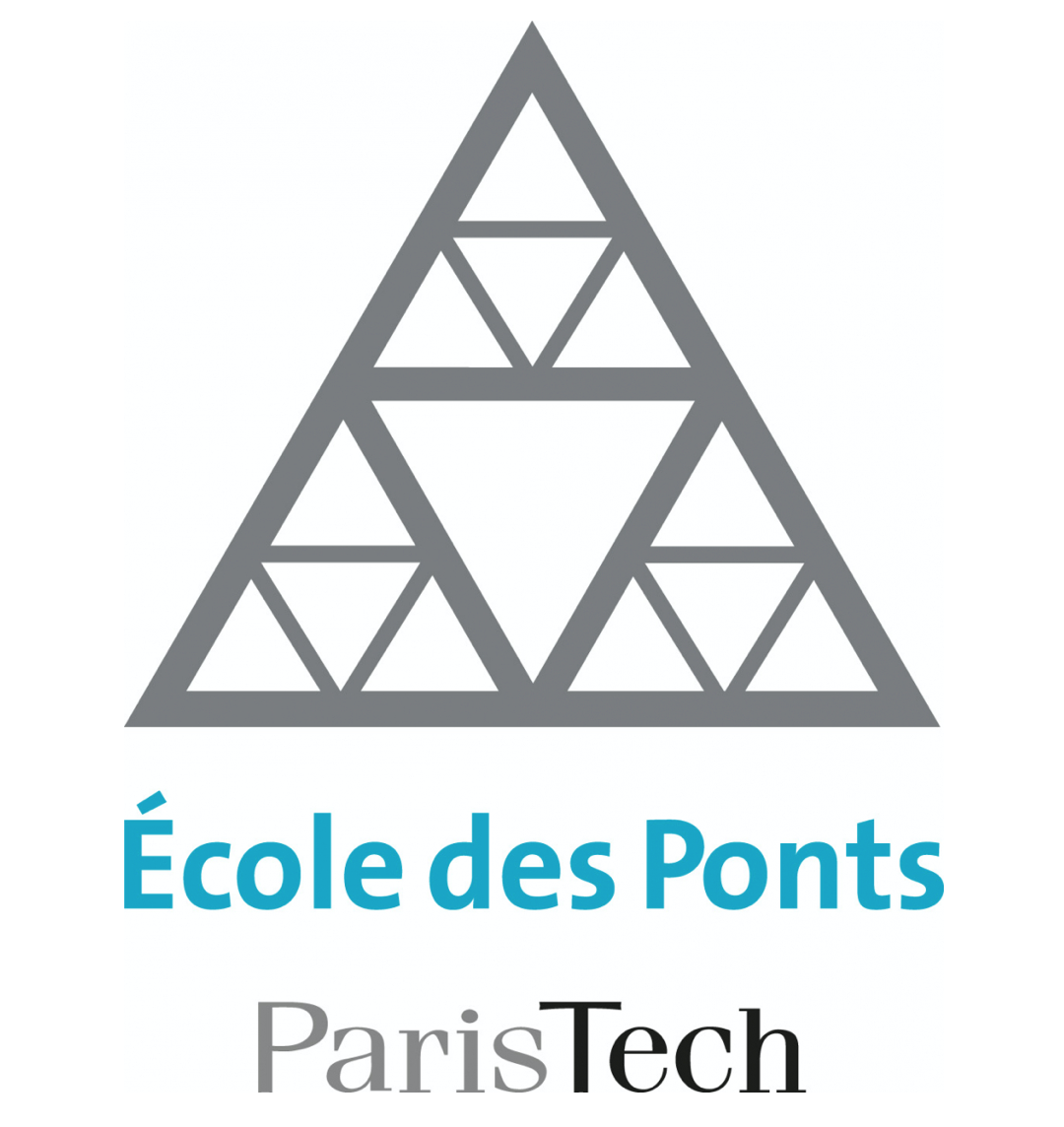} 

\end{center}	

        \vspace{0 cm}
        {\scshape\LARGE Paris School of Economics \par}
        \vspace{0 cm}
        {\scshape\LARGE École des Ponts ParisTech \par}
        \vspace{0 cm}
        {\scshape\Large 2020 - 2021 \par}
        \vspace{0 cm}
        {\scshape\Large Master Thesis\par}
        \vspace{0 cm}

        {\Large\bfseries Bayesian hierarchical analysis of a multifaceted program against extreme poverty\par}
        \vspace{0 cm}

        \vspace{0 cm}

\vspace{0 cm}
Author:\par
Louis CHARLOT \par

\vspace{0 cm}
Supervisor:\par
Prof. Luc BEHAGHEL (PSE, INRAE, J-PAL) \par

\vspace{0 cm}
Referee:\par
Prof. Philipp KETZ (PSE, CNRS) \par

\end{center}

\clearpage
\restoregeometry 
\justify

\newgeometry{bottom=2cm, top=2cm, left=2.2cm, right=2.2cm} 

\section*{Acknowledgements}

I have to start by thanking immensely Professor Luc Behaghel, my master thesis supervisor at Paris School of Economics (PSE), for the precious advice, ideas and knowledge he transmitted to me during this thesis.
Without his idea of trying to explore the possibilities that Bayesian statistics can bring to the analysis of an economic problem, I probably would not have embarked on this exciting adventure of discovery of a branch of statistics that was unknown to me.
This discovery of Bayesian statistics was for sure greatly facilitated by the diverse reading advice he gave me.
His different organization recommendations have also been very useful.
Finally, I also want to thank him for the great serenity he transmitted to me during our numerous discussions. 

I want also to warmly thank Professors Liam Wren Lewis and Jérémie Gignoux. 
The Development dissertation workshops they organized have been not only very useful for the elaboration of many ideas of my master thesis, but also extremely enriching and interesting.

I also want to thank very much Professors Abhijit Banerjee, Esther Duflo, Nathanael Goldberg, Dean Karlan, Robert Osei, William Parienté, Jeremy Shapiro, Bram Thuysbaert and Christopher Udry, who made the data of the multifaceted program totally available, without which my analysis would not have been possible.
I thank in particular Professor William Parienté for the additional information he gave me about this very interesting program.

I thank very much also Professor Rachael Meager, who made the data she used for her Bayesian analysis of microcredit expansion totally available, without which the comparison of my results with the ones she obtained in her analysis wouldn't have been possible.

I thank very much the Stan Development Team for their precious recommendations, that facilitated very much the implementation of my models.

I also want to thank Professor Pierre Jacquet and my friend Benoît Malézieux, PhD student at Inria, for the very insightful discussions we had on several topics related to this master thesis.

I finally want to thank my professors at PSE and École des Ponts, who permitted me to improve my understanding about economics and econometrics questions, which has been very useful for this thesis.

\pagebreak

\section*{Abstract}

Keywords : Multifaceted, Poverty, Bayesian, Hierarchical, Development, Economics.

\bigbreak

The evaluation of a multifaceted program against extreme poverty in different developing countries gave encouraging results, but with important heterogeneity between countries.
This master thesis proposes to study this heterogeneity with a Bayesian hierarchical analysis.
The analysis we carry out with two different hierarchical models leads to a very low amount of pooling of information between countries, indicating that this observed heterogeneity should be interpreted mostly as true heterogeneity, and not as sampling error.
We analyze the first order behavior of our hierarchical models, in order to understand what leads to this very low amount of pooling.
We try to give to this work a didactic approach, with an introduction of Bayesian analysis and an explanation of the different modeling and computational choices of our analysis.

\pagebreak

\tableofcontents
\pagebreak

\listoffigures
\pagebreak

\listoftables
\pagebreak

\newcommand{\listappendicesname}{Appendices}
\newlistof{appendices}{apc}{\listappendicesname}
\newcommand{\appendices}[1]{\addcontentsline{apc}{appendices}{#1}}
\listofappendices

\newpage
\section{Introduction}

The reduction of extreme poverty is at the hearth of development economics. 
The current COVID-19 pandemic, that might increase global poverty for the first time since 1990, poses a real challenge to the UN Sustainable Development Goal of ending poverty by 2030, according to the first estimates by Sumner et al. (2020)\cite{sumner2020}. 
This additional difficulty makes the fight against extreme poverty even more important.
While several evaluations provide evidence of the lack of efficiency of the popular solution of microcredit (Banerjee et al., 2015a\cite{banerjeeA}, Meager, 2019\cite{meager}, Bateman and Chang, 2012\cite{bateman}), some researchers propose a new "multifaceted" approach to tackle extreme poverty (Banerjee et al., 2015\cite{banerjeeC}, Bandiera et al., 2017\cite{bandiera2017}, Banerjee et al., 2018\cite{banerjee18}).

The idea of the multifaceted program is to provide a productive asset with related training and support, as well as general life skills coaching, weekly consumption support for some fixed period, access to savings accounts, and health information or services to the poorest households in a village.
The different components of this intervention are designed to complement each other in helping households to start a productive self-employment activity and exit extreme poverty.
Banerjee et al. (2015)\cite{banerjeeC} evaluate this intervention with six randomized trials carried out in six different countries (Ethiopia, Ghana, Honduras, India, Pakistan, and Peru), with a total of 10,495 participants.
If they find positive intention-to-treat (ITT) effects of the intervention for most of the outcomes when looking at all sites pooled together, it is not always the case when they analyze the impact of the intervention for each site separately. Particularly, 24 months after the start of the intervention, they find that asset ownership increases significantly in all sites but Honduras.
Given this heterogeneity of the results between the different sites, the authors conclude that it would be important to study the significant site-by-site variation in future work. 

The purpose of this master thesis is to study this heterogeneity using Bayesian statistics. 
More precisely, we use two different Bayesian hierarchical models to provide an alternative to the original frequentist analysis provided by Banerjee et al. (2015)\cite{banerjeeC}. 
Our first hierarchical model is inspired from the model proposed by Rubin (1981)\cite{rubin1981} and analyzed by the Stan Development Team (2020)\cite{stanRef2021} and Gelman et al. (2013)\cite{gelman2013}. 
It uses directly as input the coefficients and standard errors of site-level regressions, and is adapted to cases where all the data is not available.
Our second model is inspired from Gelman and Hill (2006, chapter 13)\cite{gelman2006} and Gelman et al. (2013, chapter 15)\cite{gelman2013} for the theoretical part, and from the Stan Development Team (2021, Stan Users Guide, chapter 1.13)\cite{stanRef2021} for the practical implementation.
Contrary to the first model, this second model can take as input the full dataset of outcomes, and individual- and site-level predictors. 
This permits to bring more information into the model, and possibly to improve predictions.
For both the models, we use recent recommendations of the Stan Development Team (2021)\cite{stanRef2021} to optimize the related calculations.
In addition, we explain our choices of priors.
We implement these hierarchical models with the language Stan (Stan Development Team, 2021)\cite{stanRef2021}, which is based on the No-U-Turn Sampler (NUTS), a recent improvement by Hoffman and Gelman (2014)\cite{hoffman2014} of the Hamiltonian Monte Carlo (HMC) sampling method.
We justify this choice of an HMC sampler over more ancient Markov Chain Monte Carlo (MCMC) samplers (Betancourt, 2017\cite{betancourt2017}, Haugh, 2021\cite{haugh2021}, Neal, 2011\cite{neal2011}).
We finally calculate the level of information pooling between sites to which our hierarchical modeling leads.

Our Bayesian hierarchical analysis leads to a very low amount of pooling of information between the different sites.  This gives us estimates of the ITT effect of the multifaceted program on asset ownership 24 months after the asset transfer that are very close to the ones of the simple no-pooling site-level regressions of the original approach of Banerjee et al. (2015). 
According to our different models, that lead all to similar results, the observed heterogeneity of the site-level estimates should thus be interpreted mainly as true inter-site heterogeneity.
\\Our average pooling of information between sites, around 3\%, is much lower than the values obtained by Meager (2019)\cite{meager} in her Bayesian hierarchical analysis of a microcredit expansion policy, ranging from 30\% to 50\% depending on the variable observed.
To understand whether this difference of our results with Meager's ones comes from the model or more fundamentally from the data, we apply our first Bayesian hierarchical model to her microcredit data, and find pooling averages of the same order of magnitude as those she found.
Therefore, the difference with Meager's results seems to come from the data rather than from the model.
\\Running several simulations, we then try to understand the reason of this difference.
We find that our Bayesian hierarchical model tends to have the following first order behavior.
When the the site-level estimates of the program impact are close enough to one another comparatively to their associated standard errors, the model seems to consider that the observed heterogeneity is mostly due to sampling error of the ITT effect measure in each site, and will therefore enact a high amount of pooling of information between sites.
This is the case with Meager's analysis.
On the contrary, when the site-level estimates of the program impact are far from one another (always using as metric their associated standard errors), the model seems to consider that the observed heterogeneity is mostly due to true heterogeneity of the ITT effect between sites, and will not enact a high amount of pooling of information between sites.
This is the case with our analysis.

The Bayesian hierarchical approach is already used to analyze the results of medical trials (Bautista et al., 2018\cite{bautista2018}).
In development economics, Meager (2019)\cite{meager} leads a Bayesian hierarchical analysis to analyze the results of several microcredit evaluations.
In our master thesis, we use different models from the ones we could see in our limited readings of medical literature.
We also use a different approach to the one proposed by Meager (2019).
Inspired by the work of Haugh (2021)\cite{haugh2021}, Mc Elreath (2016)\cite{McElreath2016}, and the Stan Developing Team (2021)\cite{stanRef2021}, we try to bring a didactic approach of the different choices we made for the modeling, the priors, the sampling methods, and the optimization of calculations.
We also decide to use different models in our work in order to compare and analyze their results.
Finally, we try to use recently developed Stan packages in order to present the results as visually as possible. 

The remainder of this master thesis is organized as follows.
In Section \ref{section1}, we introduce the multifaceted program and its research context.
In Section \ref{section2}, we present the utility of the Bayesian hierarchical approach for our analysis.
Then, in Section \ref{section3}, we present the hierarchical models we use, and the related optimization and prior choices.
In Section \ref{section4}, we present the results we obtain with the different models.
In Section \ref{section5}, we interpret the results and see what they bring to our knowledge of the multifaceted program.
Finally, in Section \ref{section6}, we present the ideas of future research and their related challenges.

\newpage
\section{The multifaceted program and its context}
\label{section1}

In this section, we introduce the multifaceted program and the research context that led to its implementation.
More precisely, we first briefly present the multifaceted program and the cash and skills constraints it is intended to release.
Given that the multifaceted program shares a similar aim with microcredit, we also give some comparison points with microcredit.
Then, in a second part, we try to summarize the important results of the first field evaluations of the multifaceted program, and mention some open questions about this program.

\subsection{The rationality of the multifaceted program}

\subsubsection{An answer to cash and skills constraints faced by the very poor}

The multifaceted program, a term coined by Banerjee et al. (2015)\cite{banerjeeC}, refers to an intervention providing a productive asset with related training and support, as well as general life skills coaching, weekly consumption support for some fixed period, access to savings accounts, and health information or services to the poorest households in a village.
The aim of this intervention, that has already reached in 2014 360,000 households in Bangladesh according to Bandiera et al. (2017)\cite{bandiera2017}, is to release both the cash and the skills constraints that prevent the very poor from exiting poverty by a successful business creation in developing countries.

Banerjee et al. (2018)\cite{banerjee18} argue that the multifaceted program has been developed with the idea that there should be complementarities between the program’s pieces.
The consumption support is intended to help the families during the setting up of their business, to avoid the sale or the consumption of the asset.
The training and the visits are there to help them not make elementary mistakes and stay motivated. 
The savings accounts are intended to encourage the households to save their earnings, and convert savings into future investments for the business.

As underlined by Bandiera et al. (2017)\cite{bandiera2017}, if such a program can permanently transform the lives of the poor, it would determine a causal link between the lack of capital and skills and extreme poverty in development countries.
To take the words of Banerjee (2020)\cite{banerjeeNobel}, the multifaceted program addresses a "very big question: are those in extreme poverty there because they are intrinsically unproductive, or are they just unlucky and caught in a poverty trap?"  

More precisely, the multifaceted program should permit to understand the reason why people stay poor. As highlighted by Balboni et al. (2021)\cite{balboni2021}, there are two rival theories trying to answer this question. 
The equal opportunity theory explains that differences in individual characteristics like talent or motivation make the poor choose low productivity jobs.                                                            The poverty traps theory explains on the contrary that access to opportunities depends on initial wealth and thus poor people have no choice but to work in low productivity jobs.
According to the poverty traps theory, by a sufficient asset transfer, training and support, the multifaceted program should permit the very poor households to exit poverty persistently by crossing this initial wealth barrier.
A success of the multifaceted program would therefore support the poverty traps theory.

\subsubsection{Is there really a cash constraint?}

The rationality of the multifaceted program lies on the existence of a cash constraint, that prevents the very poor to exit poverty by successful business creation.
The identification of such a cash constraint, preventing high returns and poverty exit for (potential) entrepreneurs, has been the aim of different randomized experiments.
The results of these RCTs seem to imply that such a cash constraint exists, but is not always the only one.

McKenzie and Woodruff (2008)\cite{mckenzie} identified that a release of the cash constraint could lead to high returns, with a RCT led on male-owned firms in the retail trade industry in Mexico. This experiment, by providing either cash or equipment to randomly selected enterprises, showed that an exogenous increase of capital generated large increases in profits. This return was very high (more than 70\%) for firms that report being financially constrained, which are the informal firms with less educated owners without entrepreneur parents.

However, a RCT led by Banerjee et al. (2015b)\cite{banerjeeB} has shown that releasing this cash constraint doesn't always lead to high returns. 
Their experiment showed poor effects of a group-lending microcredit in Hyderabad, which targets women who may not necessarily be entrepreneurs: the demand of credit was lower than expected (informal borrowing declined with the emergence of microcredit and there was no significant difference in the overall borrowed amount) and there was no increase of the overall consumption. Finally, business profits increased only for the ones who already had the most successful businesses before microcredit.

These examples show that a cash constraint seem to exist, but releasing it doesn't seem sufficient for all the groups of population to exit poverty. It worked for male-owned firms in the retail trade industry (McKenzie and Woodruff, 2008)\cite{mckenzie} but not for women who may not necessarily be entrepreneurs (Banerjee et al., 2015b)\cite{banerjeeB}. Other constraints seem to exist.

\subsubsection{Why not implementing microcredit instead?}

We just saw that the poor seem to face a cash constraint and we know that the aim of microcredit is precisely to release this cash constraint. So why don't we just implement microcredit instead of the multifaceted program?
If microcredit and the multifaceted program are similar by the release of a cash constraint, they differ importantly on other points.
Indeed, the multifaceted program also releases a skills constraint and it doesn't require people to reimburse what they received. 

The disappointing impact of microcredit has been established by Banerjee et al. (2015a)\cite{banerjeeA}, who aggregated the evidence of 6 RCTs about microcredit expansion led in very different contexts: 6 countries, urban and rural areas, different borrower characteristics (women only or both genders), loan characteristics (from 12\% to 110\% interest rates, individual or group lending), and lender characteristics (NGO or for profit).
Indeed, they found a lack of evidence of transformative effects of microcredit expansion on the average borrower: microcredit does not reduce poverty, improve living standards or social indicators on average. However, businesses tend to expand, but this expansion does not lead to better standards of living as it consists of an equal value replacement of wage income by business income. 
Further analysis by Meager (2019)\cite{meager} shows that, if microcredit has not significant effect on the average borrower, it can have positive but highly variable effects on profits for experienced people who already owned a business.
\\This confirms the fact that microcredit is not a sufficient solution for the very poor, who have no business experience.

In addition, the general equilibrium effects of microcredit have been questioned by some economists. 
For instance, Bateman and Chang (2012)\cite{bateman} claim that microcredit can harm a developing country economy by several mechanisms. These different mechanisms, that may affect negatively the economy at a large scale, cannot be observed during an RCT evaluation. First, through its high interest rates and short maturities loans, microcredit model tends to encourage the development of unsophisticated micro-enterprises (retail and service operations) instead of longer-term returns growth-oriented enterprises that use sophisticated technologies. This can be problematic as technically innovative ideas and institutions can be important actors of development. 
Second, as argued by Karnani (2007)\cite{karnani}, the microcredit model ignores the importance of scale economies by supporting the development of tiny micro-enterprises at the expense of large, productive and labor-intensive industries. In addition, by increasing the number of micro-enterprises without increasing substantially the demand, microcredit tends to increase competition and decrease prices thus income for the existing enterprises on the market. 

The disappointing impact of microcredit for the very poor and its potentially harmful general equilibrium effects justify the search of another policy, like the multifaceted program.

\subsection{Current knowledge about the impact of the multifaceted program}

\subsubsection{Encouraging results from the first evaluations}
\label{autrePourPrior}
We saw previously that classical microcredit does not significantly improve lives of the very poor.
Besides releasing more the cash constraint for the very poor, multifaceted programs present other advantages compared to microcredit.
Indeed, as households do not have to pay back the asset transfer, they will probably take more risks by investing more in their new activity. In addition, the training makes them gain some "entrepreneurial experience" that was maybe missing in the microcredit experiments.
\\Two similar forms of the multifaceted program have been analyzed for now. 

The \textbf{first form} of multifaceted program has been evaluated by Bandiera et al. (2017)\cite{bandiera2017} with a randomized control trial covering over 21,000 households in 1,309 villages in Bangladesh.
This program has been implemented at a large scale (400,000 women in 2011) by the Bangladesh Rural Advancement Committee (that began its activities with classical microcredit in 1974) and consists of a sizable transfer of productive assets and skills to the poorest women of rural villages.
More precisely, eligible women are offered a set of possible business activities, like livestock rearing or small retail operations, coupled with complementary and intensive training in running whichever business activity they chose. 
The program also provides a subsistence allowance to eligible women for the first 40 weeks after the asset transfer, delivers health support and training on legal, social and political rights, organizes committees made up of village elites which offer support to program recipients, and encourages saving with BRAC during the program and borrowing from BRAC microfinance at the end of the program.
Thus, this program significantly relaxes both capital constraints (value of the asset transfer is worth roughly ten times baseline livestock wealth) and skills constraints (the value of the two-year training and assistance which women receive is of a similar magnitude).
\\ To measure long term effects of the program, households were surveyed 2, 4 and 7 years after the program implementation.
Bandiera et al. (2017) showed that this intervention enabled the poorest women to shift out of agricultural labor by running small businesses. This shift, which persists and strengthens after assistance is withdrawn, leads to 21\% higher earnings than their counterparts in control villages.
There is an increase of self-employment and a decrease of wage-employment for treated women, who work more regularly and are more satisfied. 
These results are more important 4 years after than 2 years after the program beginning, and sustained after 7 years, showing that these positive effects seem to last, guaranteeing a sustainable path out of poverty. A quantile effect analysis shows that the program has a positive effect on earnings and expenditures at all deciles, but that these effects are slightly better for high deciles.
\\ However, the question of external validity is still open: the authors mention that a similar program in West-Bengal gave good results, but that it wasn't the case in Andhra Pradesh. 
They explain this failure by the fact that the Government of Andhra Pradesh simultaneously introduced a guaranteed-employment scheme that substantially increased earnings and expenditures for wage laborers.

A \textbf{second form} of multifaceted program (the one we will study) has been evaluated by Banerjee et al. (2015)\cite{banerjeeC} with 6 RCTs in 6 different countries (Ethiopia, Ghana, Honduras, India, Pakistan and Peru), with a total of 10,495 participants.
This program is slightly different from the one of Bandiera et al. (2017): the focus is always on ultra poor, but this time also men are targeted. 
However, the aim of the program is still to help very poor households to start or continue with a self-employment activity through a release of cash and skills constraints. 
More precisely, it is this time a combination of a productive asset transfer, ranging from raising livestock to petty trade, of technical skills training on managing the particular productive asset and high-frequency home visits, but also of consumption support (regular transfer of food or cash for a few months to about one year), saving support (access to a savings account and in some instances a deposit collection service and/or mandatory savings) and some health education, basic health services, and/or life-skills training.
There are some differences from one site to another: for instance, only 4 sites partnered with microfinance institutions able to provide access to savings accounts (that were more or less compulsory according to sites).
\\ They also find very positive results for their program: consumption, revenue and income of the treated very poor households increase, and this positive effects persist for at least one year after the program ends. 
This increase, though significant at all tested quantiles, is lower at the bottom of the distribution. Although results vary across countries, the general pattern of positive effects that persist for at least one year after the end of the program is common across all countries, with weaker impacts in Honduras and Peru.
In addition, if we consider total consumption as the measure for benefits, all the programs except Honduras one have benefits greater than their costs (from 133\% in Ghana to 433\% in India).

So, in both cases the multifaceted program seems to better address the issue of extreme poverty exit than what we saw with microcredit.
\\ In addition, the multifaceted program has some advantages compared to a classical unconditional cash transfer of the same cost according to Bandiera et al. (2017)\cite{bandiera2017}: they showed that this program gives higher earnings on average, and permits households to smooth more naturally their consumption.

We can also mention that a recent paper by Banerjee et al. (2018)\cite{banerjee18} mentioned that a seven-year follow up led in Bangladesh and India showed that impacts persisted for these countries.
Even if this follow up has not been carried out in all the countries, it gives some hopes of long-lasting positive effects of the program.

Finally, Banerjee et al. (2018)\cite{banerjee18} find also that both a savings-only treatment and an asset-only treatment have worse effects than the full multifaceted program.
More precisely, the savings-only program has much weaker effects on consumption one year after the end of the program, while the asset-only treatment has no evidence of any positive welfare effects.
These first results provide some evidence of the complementarity of the asset transfer with the other components of the program.

\subsubsection{Still some open questions about impact heterogeneity and general equilibrium effects}
\label{openQ}
There are still some open questions about the impact of the multifaceted program.

Let us first note that some of the critics of microcredit by Bateman and Chang (2012)\cite{bateman} can be also applied to this multifaceted intervention. Indeed, as the program ultimate goal is to increase consumption and life standards of the very poor by an occupational shift towards entrepreneurship, we can question if this multiplication of tiny micro-enterprises can harm the economy of a country. It is not easy to answer to these critics as this program has been evaluated for now only through RCTs at medium-term and medium-scale.
\\ Nevertheless, as the main goal of this program is to extract the very poor from the poverty trap (no capital, no skills), and not necessarily to improve the economic development of an entire country (which was the initial claim of microcredit), these uncertainties do not really question the multifaceted program.
\\ Furthermore, the first results obtained by Banerjee et al. (2015)\cite{banerjeeC} from the 3 sites where randomization allowed the examination of spillover effects show no spillover effects on primary economic outcomes such as consumption and income, and no significant spillover effects at the 5\% level on any variable after accounting for multiple hypothesis testing.
Bandiera et al. (2017)\cite{bandiera2017} found for their similar program that the gains of treated households did not crowd out the livestock businesses of non-eligible households while the wages these receive for casual jobs increase as the poor reduce their labor supply.
Given the fact that implementing the program at larger scale will mainly require increasing geographic coverage, rather than increasing the proportion of households reached in each village, we can hope that this absence of direct negative spillover effects keeps being true even if the program is scaled up.

Another open question, highlighted by Banerjee et al. (2015)\cite{banerjeeC}, concerns the heterogeneity of the impact in the different locations.
Indeed, if the authors find positive intention-to-treat (ITT) effects of the intervention for most of the outcomes when looking at all sites pooled together, it is not always the case when they analyze the impact of the intervention for each site separately. Particularly, 24 months after the start of the intervention, they find that asset ownership increases significantly in all sites but Honduras. Given this heterogeneity of the results between the different sites, the authors conclude that it would be important to study the significant site-by-site variation in future work. 
The purpose of this master thesis is to study this heterogeneity using Bayesian statistics.

\newpage
\section{Utility of the Bayesian hierarchical approach}
\label{section2}

\subsection{The limits of the original non-hierarchical analysis}

The data of Banerjee et al. (2015)\cite{banerjeeC} is hierarchical by nature: we look at households characteristics, and these households are located into different countries.
With this type of data, three different approaches are possible.
The first one is to treat the households of different countries separately, supposing that what we learn from a country does not provide any information about another country: this is the no-pooling approach.
The second one is to treat the households of different countries without any distinction, supposing that what we observe in a country tells us exactly what happens in another country: this is the full-pooling approach.
We suppose here that all experiments might be estimating the same quantity.
The third approach, which is the one we are interested in, is to implement partial pooling at a level based on what we observe: this approach is halfway between the two previous ones.

In the original study by Banerjee et al. (2015)\cite{banerjeeC}, results are presented with the two first approaches: the authors give both the results for each country separately and the results obtained with the data of all countries pooled together.
These two approaches give interesting results regarding the households asset ownership evolution at the end of the program (24 months after the asset transfer): if the full-pooling approach shows a global significant increase of asset ownership, the no-pooling approaches shows that this increase is significant in all sites but Honduras.

As highlighted by Gelman et al. (2013)\cite{gelman2013} and Mc Elreath (2016)\cite{McElreath2016}, neither of these two extreme approaches (that is the separate analyses that consider each country separately, and the alternative view of a single common effect that leads to the pooled estimate) is intuitive.
Indeed, the full-pooling approach ignores possible variations between sites.
In our case, it would imply that we believe that the probability that the true effect of the multifaceted program on asset ownership in Honduras (where no significant effect of the multifaceted program is measured) being lower than in Ghana (where a significant positive effect is measured) is only 50\%.
On the contrary, considering each country separately would imply that we believe that the large positive and significant effects of the multifaceted program measured in other countries do not provide any hope for the true effect in Honduras being greater than what was measured.
We can see that neither of these approaches is fully satisfactory: it would be interesting to have a compromise combining information from all countries without assuming all the true ITT effects to be equal.
A hierarchical model, that takes into account the two levels of the data (household level and country level) and that we use in Section \ref{section3}, provides exactly this compromise.

\subsection{The rationality of the Bayesian approach}

In this master thesis, we use the Bayesian approach to implement our hierarchical models. 
Before highlighting the advantages of this approach over the frequentist approach, we give in this section a very quick introduction of Bayesian statistics.

As explained by Haugh (2021)\cite{haugh2021}, Bayesian statistics are based on the following application of the Bayes theorem:
\[
\pi(\theta \mid y)
=\frac{p(y \mid \theta) \cdot \pi(\theta)} {p(y)}
=\frac{p(y \mid \theta) \cdot \pi(\theta)} {\int_{\theta} p(y \mid \theta) \cdot \pi(\theta)}
\]
with $\theta$ being an unknown random parameter vector,
$y$ a vector of observed data with likelihood $p(y \mid \theta)$,
$\pi(\theta)$ being the \textbf{prior distribution} of $\theta$, that is the distribution we assume for $\theta$ before observing the data,
$\pi(\theta \mid y)$  being the \textbf{posterior distribution} of $\theta$, that is the updated distribution of $\theta$ after observing the data.
This formula is at the basis of Bayesian statistics: it provides a rule to update probabilities when new information appears. 
More precisely, the parameters of interest $\theta$ are assumed random and the observation of new data $y$ permits to update our prior beliefs about the distribution of these parameters of interest.

Contrary to the \textbf{frequentist} approach where the parameter vector $\theta$ is taken fixed and the dataset $y$ is taken with uncertainty, the \textbf{Bayesian} approach fixes the dataset $y$ but takes the parameter vector $\theta$ with uncertainty.

As highlighted by Haugh (2021)\cite{haugh2021}, the selection of the prior is an important element of Bayesian modeling.
With a few amount of data, the influence of the prior choice on the final posterior can be very large. It can be in this case important to understand the sensitivity of the posterior to the prior choice.
An ideal prior should capture the relevant information we have before observing the data, but should be dominated by the data when there is plenty of data.
In Section \ref{section3}, we will justify the different choices of priors that we use for our hierarchical models.

\subsection{The advantages and drawbacks of the Bayesian approach}

There are several advantages of the Bayesian approach over the frequentist one, as highlighted by Haugh (2021)\cite{haugh2021}, Gelman (2013)\cite{gelman2013}, and Thompson and Semma (2020)\cite{thompsonSemma2020}.

First of all, the choice of priors permits to express our prior beliefs about the quantities of interest: we will see several examples on this point with our hierarchical models in Section \ref{section3}.
Another very useful advantage of Bayesian statistics in the case of our hierarchical analysis is the possibility to build very flexible models.
The fact that we can visualize the final distributions with Bayesian statistics also leads to results easy to understand and interpret. 

The more intuitive Bayesian interpretation of the results is particularly visible with \textbf{credible intervals}, the Bayesian version of frequentist confidence intervals.
While the 95\% credible interval for a parameter $\theta$ can be interpreted as having a 95\% probability of containing $\theta$, this is not true for the 95\% confidence interval for $\theta$.
Indeed, a frequentist analysis assumes that $\theta$ is unknown but not random.
Therefore, a 95\% confidence interval contains $\theta$ with probability 0 or 1.
The interpretation of confidence interval is less intuitive: if we repeat the experiment an infinity of times, we expect the 95\% confidence interval to contain the true value of $\theta$ 95\% of the time.

As mentioned by Haugh (2021)\cite{haugh2021}, there are also some drawbacks with the Bayesian approach.
The first one is the subjectivity induced by the choice of the prior: we will discuss this difficulty and how to partially handle it in our practical implementation.
The second one is the high computational cost of Bayesian analysis: it is in particular the case of the Bayesian sampling challenge, that we will address in the next two sections.

\subsection{Markov Chain Monte Carlo (MCMC) to solve the sampling problem of the Bayesian approach}

An important challenge in Bayesian statistics is the resolution of the sampling problem. 
As highlighted by Haugh (2021)\cite{haugh2021}, it is the recent progress in sampling methods that permitted to do the calculations necessary to Bayesian analysis. 
After a century of domination of frequentist analysis, this sampling progress has brought Bayesian analysis back to the forefront. 

The \textbf{sampling problem} is in fact the problem of simulating from the posterior $\pi(\theta \mid y)$ without knowing its denominator in the expression:
\[
\pi(\theta \mid y)
=\frac{p(y \mid \theta) \cdot \pi(\theta)} {\int_{\theta} p(y \mid \theta) \cdot \pi(\theta)}
\myeq \frac{\Tilde{p}(\theta)}{Z_p}
\]
This problem appears in Bayesian models where 
$\Tilde{p}(\theta) \myeq p(y \mid \theta) \cdot \pi(\theta)$ is easy to compute but 
$Z_p \myeq {\int_{\theta} p(y \mid \theta) \cdot \pi(\theta) d \theta}$ is very difficult or impossible to compute.

Several Markov Chain Monte Carlo (\textbf{MCMC}) sampling methods have been developed to solve this sampling problem (Betancourt, 2017\cite{betancourt2017}, Haugh, 2021\cite{haugh2021}).
The common idea of these methods is to construct a Markov chain for which the distribution we want to sample (that is $\frac{\Tilde{p}(\theta)}{Z_p}$) is the stationary distribution.
After having achieved this construction, we only need to simulate the Markov chain until stationarity is achieved to get the desired sample.

\subsection{Our choice of Hamiltonian Monte Carlo (HMC) and Stan}

In order to obtain a high quality and quick sampling for our hierarchical models, we use the Hamiltonian Monte Carlo (\textbf{HMC}) version of MCMC sampling, as recommended by Neal (2011)\cite{neal2011}.
HMC is a recent version of the MCMC sampling that uses the formalism of Hamiltonian dynamics in order to improve the quality and the efficiency of sampling for multidimensional and complex models. 

Very shortly, Hamiltonian dynamics is a discipline of physics that permits to describe the movement of a physical particle thanks to Hamiltonian formalism (a particular way to write the physical equations of movement).
The very interesting point for us is that the formalism of Hamiltonian dynamics can be applied to very diverse systems, including sampling in Bayesian statistics.
The idea is to see the posterior distribution we want to sample as the position $q$ of a particle in Hamiltonian dynamics. We then associate randomly a distribution to the momentum $p$ (mass multiplied by velocity) of this same particle, and simulate the evolution of this particle in the state space ($p$, $q$) thanks to the leapfrog method (and improvement of the Euler method that is more adapted to Hamiltonian dynamics) and Hamiltonian dynamics.
We finally accept the new state ($p$, $q$) of the particle with a probability that increases with the energy of the particle at this new state.
If we do not accept the new state, we stay at the previous state.
As this very short introduction of this method provides only the basic idea of HMC method, we recommend to the readers interested by these questions the very interesting paper by Neal (2011)\cite{neal2011}, that goes more into details in a very didactic way.

The original HMC algorithm requires the user to set the step size $\epsilon$ and the number of steps $L$ for the leapfrog method part of the algorithm.
Hopefully, Hoffman and Gelman (2014)\cite{hoffman2014} proposed a new version of HMC that sets automatically $L$ and $\epsilon$, which is called the No-U-Turn Sampler (\textbf{NUTS}). 

Given the advantages of HMC sampling for hierarchical models, we use in this master thesis the probabilistic programming language Stan, presented by Carpenter et al. (2017)\cite{carpenter2017} and the Stan Development Team (2021)\cite{stanRef2021}, that provides Bayesian inference with NUTS.

\newpage
\section{Two hierarchical models for the multifaceted program} 

\label{section3}
We will now present the two models we used to carry out our hierarchical analysis of the impact of the multifaceted program on asset ownership.
While our first model uses directly as inputs the coefficients and standard errors of the site-level regressions, our second model uses the full dataset of individuals as input.
In this section, we will present both models, their optimization and the related prior choices. We will also see the advantages and difficulties of each one.

\subsection{A first model using the results of the site-level regressions}
\label{FirstMod}

\subsubsection{The model}   

Our first hierarchical model is inspired from the model proposed by Rubin (1981)\cite{rubin1981}, commented by the Stan Development Team (2020)\cite{RStanCRAN}, Gelman et al. (2013, chapters 5 and 13)\cite{gelman2013}, and Thompson and Semma (2020)\cite{thompsonSemma2020}.
In this model, we begin by carrying out the following regression in each site $s$:
\[
Y_{i} = \alpha_s + \tau_s T_i + \epsilon_i
\]
with $Y_{i}$ outcome of interest of individual $i$, $T_i$ the intention-to-treat (ITT) indicator for individual $i$, and $\tau_s$ the associated ITT effect.

\begin{figure}
    \centering
    \caption{Visual summary of the first model}
    \label{fig:dessin1}
    \begin{forest}
  for tree={
    math content,
  },
  delay={
    where content={}{
      if level=0{}{
        content=\ldots,
        math content,
        no edge,
        fit=band,
      },
    }{
      rounded corners,
      outer color=blue!20,
      inner color=blue!15,
      minimum height=1cm,
      minimum width=1cm,
      draw,
      drop shadow,
    },
  },
  label levels,
  [, phantom
    [$\tau$, plain content, level label=Universe
      [\tau_1 \sim N(\tau \text{,}\ \sigma^2), level label=Site-level
        [\hat{\tau}_1 \sim N(\tau_1 \text{,}\ \hat{\sigma}_1^2), level label=Measure]
      ]
      []
      [\tau_6 \sim N(\tau \text{,}\ \sigma^2)
        [\hat{\tau}_6 \sim N(\tau_6 \text{,}\ \hat{\sigma}_6^2)]
      ]
    ]
  ]
\end{forest}
\end{figure}
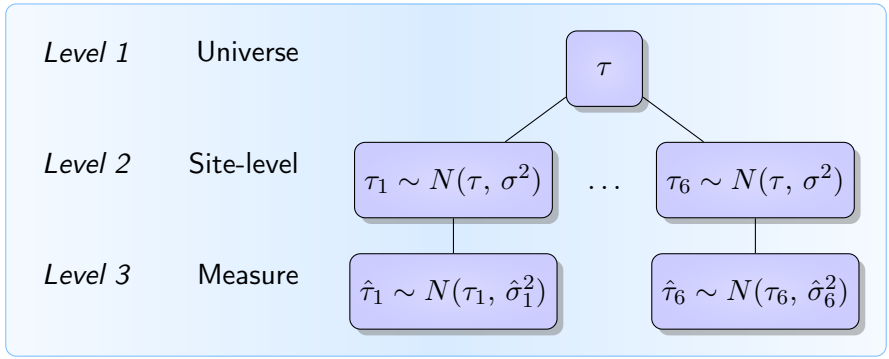

We then use the the estimates $\Hat{\tau}_s$ of the treatment effect $\tau_s$
and their associated standard errors $\Hat{\sigma}_s$ obtained with each site-level regression to build the following model:
\[
\Hat{\tau}_s \sim N(\tau_s,\Hat{\sigma}_s^2)
\]
\[
\tau_s \sim N(\tau,\sigma^2)
\]
In other words, we assume here that the observed treatment effect $\Hat{\tau}_s$ for site $s$ is drawn from the normal distribution $N(\tau_s,\Hat{\sigma}_s^2)$ of the true treatment effect for site $s$. 
According to DuMouchel (1994)\cite{dumouchel1994}, if the estimates of site treatment effects are not based on very small samples, this assumption is likely to be a good one.
We also assume here that the mean $\tau_s$ of this  normal distribution of the treatment effect for site $s$ is drawn from the normal distribution $N(\tau,\sigma^2)$ of the treatment effect over all possible sites in the universe.
$\tau$ and $\sigma$ are considered as random and they must be assigned a prior.
As we believe that no prior knowledge exists about $\tau$ and $\sigma$, we assign them weakly informative priors:
\[
\tau \sim N(0,5) \text{ and } \sigma \sim halfCauchy(0,5)
\]
The choice of a half-Cauchy distribution (that is a Cauchy distribution defined over the positive reals only) is usual for a parameter, like $\sigma$, that is strictly positive, as highlighted by Mc Elreath (2016)\cite{McElreath2016}.
We give a visual summary of this first model in Figure \ref{fig:dessin1} for more intuition.

Let us note that this model is the Bayesian version of the frequentist random effect model for meta-analysis, as highlighted by Thompson and Semma (2020)\cite{thompsonSemma2020} and Biggerstaff et al. (1994)\cite{biggerstaff1994}.
In the frequentist version, used for instance by Fabregas et al. (2019)\cite{fabregas2019} in development economics, the difference is that parameters $\tau$ and $\sigma$ are not considered as random and are directly estimated.

\subsubsection{Optimization for HMC computations}
\label{firstreparam}
As highlighted by Gelman et al. (2013)\cite{gelman2013}, HMC can be very slow to converge with hierarchical models. This issue can be solved with a reparametrization of the model.
More practically, the original model
\[
\Hat{\tau}_s \sim N(\tau_s,\Hat{\sigma}_s^2)
\]
\[
\tau_s \sim N(\tau,\sigma^2)
\]
can lead to a very slow convergence for HMC because no single step size works well for the whole joint distribution of $\tau_s$, $\tau$ and $\sigma$.
Indeed, according to Gelman et al., the trajectories in the HMC are unlikely to go in the region where $\sigma$ is close to 0 and then are unlikely to leave this region when they are inside it.
Therefore, they propose to write this model with the following parametrization:
\[
\Hat{\tau}_s \sim N(\tau + \sigma \eta_s,\Hat{\sigma}_s^2)
\]
\[
\eta_s \sim N(0,1).
\]
The idea is here to take the means and standard deviations out of the original Gaussian distribution, which leaves only a standardized Gaussian prior.
If the model is technically the same as before, this "non-centered" form of the model permits to sample much more efficiently according to Mc Elreath (2016)\cite{McElreath2016}.

\subsection{A second model using both site-level and household-level information}
\label{secondmodelall}

\subsubsection{The model}
\label{deuxmod}

We also carry out the data analysis with a second hierarchical model, inspired from Gelman and Hill (2006, chapter 13)\cite{gelman2006} and Gelman et al. (2013, chapter 15)\cite{gelman2013} for the theoretical part, and from the Stan Development Team (2021, Stan Users Guide, chapter 1.13)\cite{stanRef2021} for the practical implementation. 
This second model belongs to the family of hierarchical regression models, recommended by Gelman et al. (2013, chapter 15)\cite{gelman2013} in the cases where there are predictors at different levels of variation (individual level and site level in our case).
Contrary to the first model of Section \ref{FirstMod}, that took as inputs only the coefficients and standard errors of site-level regressions $\Hat{\tau}_s$ and $\Hat{\sigma}_s$, this second model can take as input the full dataset of outcomes, and individual- and site-level predictors.
This permits to bring more information into the model, and possibly to improve predictions.

The main idea of this second hierarchical model is to suppose that the outcome of interest $y_i$ for an individual $i$ has a distribution of the form:
\[
y_{i} \sim N\left( X_{i} \beta_{site[i]}\ ,\ \sigma_{site[i]}^2 \right)
\]
with $X_{i}$ being the vector of individual-level predictors and $site[i]$ being the site where individual $i$ is located (including an intercept through $X_{i,1} = 1$).
We give a weakly-informative prior to $\sigma_s$, as recommended by the Stan Development Team (2021).

For instance, in the case of the multifaceted program, we could take a vector $X_{i} = (1, T_{i})$, with $T_{i}$ the ITT dummy for individual $i$.
By renaming $\beta_{site[i]}$ as $(\alpha_{site[i]}, \tau_{site[i]})$, this would lead to the model:
\[
y_{i} \sim N(\alpha_{site[i]} + \tau_{site[i]} \cdot T_{i}\ ,\ \sigma_{site[i]}^2)
\]
We can notice that the intercept $\alpha_{site[i]}$, the ITT effect $\tau_{site[i]}$ and the variance $\sigma_{site[i]}^2$ all vary with the site.
Another choice could be to take for instance $X_{i} = (1, T_{i}, y_{baseline,i})$ with $y_{baseline,i}$ being the baseline value of the outcome of interest for individual $i$.

Returning to the general case and following Gelman and Hill (2006)\cite{gelman2006}, we suppose that the coefficients $\beta_s$ for each site $s$ follow a multivariate normal distribution:
\[
\beta_s \sim N ( \mu_s, \Sigma )
\]
For instance, in the case where $X_{i} = (1, T_{i})$, we would have:
\[
\beta_s 
= 
\begin{pmatrix}
  \alpha_s \\ 
  \tau_s \\ 
\end{pmatrix}
\sim
N\left(
\begin{pmatrix}
  \mu_{\alpha,s} \\ 
  \mu_{\tau,s} \\ 
\end{pmatrix}
\ ,\
\Sigma
\right),
\text{ with }
\Sigma
=
\begin{pmatrix}
  \sigma_{\alpha}^2 & \rho \sigma_{\alpha} \sigma_{\tau} \\ 
  \rho \sigma_{\alpha} \sigma_{\tau} & \sigma_{\tau}^2  \\ 
\end{pmatrix}
\text{and $\rho$ correlation parameter.}
\]
In their model, Gelman and Hill (2006)\cite{gelman2006} take for the prior mean $\mu_s$ of $\beta_s$ a simple vector parameter $\mu$, which does not vary across sites.
As recommended by the Stan Development Team (2021)\cite{stanRef2021}, we choose instead to include in our model site-level information through site-level predictors $Z_s$ (including an intercept through $Z_{s,1} = 1$).
The idea is to model the prior mean $\mu_s$ of $\beta_s$ itself as a regression over the site-level predictors $Z_s$ as follows:
\[
\beta_s \sim N( Z_{s} \gamma , \Sigma )
\]
with $\gamma$ being the vector of site-level coefficients. 
We can give to each element of $\gamma$ a weakly informative prior, such as:
\[
\gamma_k \sim N(0,5). 
\]

Regarding the prior on the covariance matrix $\Sigma$ of $\beta_s$, Gelman and Hill (2006)\cite{gelman2006} propose to use a scaled inverse Wishart distribution. This choice is motivated by the fact that the inverse Wishart distribution is the conjugate prior (prior with the same distribution family as the posterior) for the covariance matrix of a multivariate normal distribution ($\Sigma$ in our case).
Indeed, using the conjugate prior is computationally convenient when using Bugs, the programming language based on the Gibbs sampler (an improvement of the first version of MCMC algorithm) used by Gelman and Hill.
As we are using Stan, which is based on Hamiltonian Monte Carlo, there is no such restriction in our case. 
Therefore, we follow instead a more intuitive approach recommended by the Stan Development Team (2021)\cite{stanRef2021}.
The idea is to decompose the prior on the covariance matrix $\Sigma$ into a scale $diag(\theta)$ and a correlation matrix $\Omega$ as follows:
\[
\Sigma = diag(\theta) \times \Omega \times diag(\theta)
\text{, with } 
\theta_k = \sqrt{\Sigma_{k,k}}
\text{ and } 
\Omega_{k,l} = \frac{\Sigma_{k,l}}{\theta_k \theta_l}.
\]
The advantage of this decomposition is that we can then impose a separate prior on scale and on correlation.
For the elements of the scale vector $\theta$, the Stan Development Team recommends to use a weakly informative prior like a half-Cauchy distribution (that is a Cauchy distribution defined over the positive reals only) with a small scale, such as:
\[
\theta_k \sim Cauchy(0, 2.5) \text{,\ \ \ constrained by } \theta_k > 0
\]
As explained by Mc Elreath (2016)\cite{McElreath2016}, both the Gaussian prior for $\gamma_k$ and the Cauchy prior for $\theta_k$ contain very gradual downhill slopes. This explains the fact that they are quickly overcome by the likelihood when the number of observations increases.
The choice of a half-Cauchy distribution is usual for a parameter, like $\theta_k$, that is strictly positive.

Regarding the correlation matrix $\Omega$, the Stan Development Team recommends to use a LKJ prior (Lewandowski, Kurowicka and Joe, 2009)\cite{LKJ2009} with parameter $\eta \geq 1$:
\[
\Omega \sim LKJcorr(\eta)
\]
The idea of a LKJ prior is to define a distribution over correlation matrices.
As explained by McElreath (2016, chapter 13.1)\cite{McElreath2016}, the parameter $\eta$ defines how skeptical the prior is of large correlations in the matrix. With a $LKJcorr(1)$, the prior is flat over all valid correlation matrices. 
With $\eta \geq 1$, extreme correlations (close to -1 and 1) are less likely.
Therefore, by setting the parameter $\eta$, we define our prior beliefs about the strength of the correlations between the coefficients of $\beta_s$.

\begin{figure}
    \centering
    \caption{Visual summary of the second model}
    \label{fig:dessin2}
    \begin{forest}
  for tree={
    math content,
  },
  delay={
    where content={}{
      if level=0{}{
        content=\ldots,
        math content,
        no edge,
        fit=band,
      },
    }{
      rounded corners,
      outer color=blue!20,
      inner color=blue!15,
      minimum height=1cm,
      minimum width=1cm,
      draw,
      drop shadow,
    },
  },
  label levels,
  [, phantom
    [$y_{i} \sim N\left( X_{i} \beta_{\text{site[i]}} \text{,}\ \ \sigma^2_{\text{site[i]}}\right)$, plain content, level label=Household-level  predictors
      [\beta_1 \sim N\left( Z_{1} \gamma \text{,}\ \ \Sigma \right), level label=Site-level predictors
        [\Sigma \text{=}\ diag(\theta) \times \Omega \times diag(\theta)]
      ]
      []
      [\beta_6 \sim N\left( Z_{6} \gamma \text{,}\ \ \Sigma \right)
      ]
    ]
  ]
\end{forest}
\end{figure}
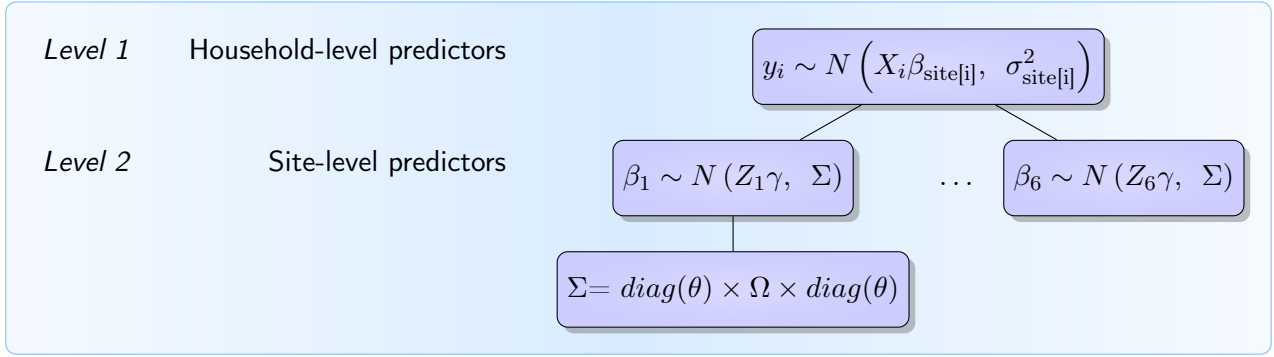

To visualize these last steps more easily, let us return to the example where $X_{i} = (1, T_{i})$.
In this case, we have:
\[
\Sigma 
=
diag(\theta) \times \Omega \times diag(\theta)
=
\begin{pmatrix}
  \sigma_{\alpha} & 0 \\ 
  0 & \sigma_{\tau}  \\ 
\end{pmatrix}
\times
\begin{pmatrix}
  1 & \rho \\ 
  \rho & 1 \\ 
\end{pmatrix}
\times
\begin{pmatrix}
  \sigma_{\alpha} & 0 \\ 
  0 & \sigma_{\tau}  \\ 
\end{pmatrix}
\]
With a $LKJcorr(1)$ prior, all values of $\rho$ (between -1 and 1) are equally possible. When we increase $\eta$ ($\eta \geq 1$), the extreme values of $\rho$ (close to -1 and 1) become less likely. 
Therefore, as our prior belief is that there should not be strong correlation between $\alpha$ and $\tau$, we take a $\eta \geq 1$, like $\eta = 2$.
We give a visual summary of this second model in Figure \ref{fig:dessin2} for more intuition.

The \textbf{advantage of this second model} compared to the first one of Section \ref{FirstMod} is that we can include both individual-level and group-level information in our model to improve predictions. 
In addition, we saw that we can include new forms of prior information if available (about correlations, standard deviations, etc.). 
However, the calculations for this second model are more heavy than for the first one, as we use the whole data as input, and not only the coefficients of site-level regressions.
The heaviness of the calculations and the more important complexity of this model can lead to very long computing times that can be an obstacle to practical implementation. 
We will now explore how to address this computational issue.

\subsubsection{Optimization for HMC computations}
\label{optideux}

To answer to the computational issue highlighted in the previous section, the Stan Development Team (2021)\cite{stanRef2021} recommends to vectorize the Stan code of our model. 
The idea is to  create local variables for $X\beta$ and $\Sigma$ that permit reducing the sampling time for $y$ and $\beta$ by avoiding unnecessary repetitions of calculations that would happen with loops (see Stan code in Appendix for more details). 

In addition, as the vectorization can be insufficient to optimize HMC sampling, the Stan Development Team proposes to combine it with a Cholesky-factor optimization.
Indeed, we can notice that, as a correlation matrix, $\Omega$ is symmetric definite positive (SDP).
Therefore, according to the Cholesky factorization theorem (van de Geijn, 2011)\cite{vandeGeijn2011}, there is a lower triangular matrix $\Omega_L$ such that:
\[
\Omega = \Omega_L \times \Omega_L^T 
\]
The idea of the Cholesky-factor optimization is to take advantage of this factorization of $\Omega$ to reduce the number of matrix multiplications in our sampling.
This can be done by defining our $\beta$ by the alternative form:
\[
\beta =  Z \gamma + \theta \times \Omega_L \times u
\]
with u being a random vector $u$ of components
\[
u_k \sim N(0,1)
\]
and $\Omega_L$ the Cholesky factor of $\Omega$.
If we want our $\Omega$ to have a $LKJcorr(\eta)$ prior, we have to assign to $\Omega_L$ the equivalent Cholesky factorized prior:
\[
\Omega_L \sim LKJcorrCholesky(\eta).
\]
We can check that we have, as desired, that
\[E(\beta)
= Z \gamma
= E(Z \gamma + \theta \times \Omega_L \times u) 
\text{, as }
u_k \sim N(0,1)
\]
and
\begin{align*}
V(\beta)
& = E\left( (\theta \times \Omega_L \times u) \times (\theta \times \Omega_L \times u)^T \right)\\
& = E\left( (\theta \times \Omega_L ) \times u \times u^T \times (\theta \times \Omega_L)^T \right)\\
& = (\theta \times \Omega_L ) \times 
E\left(u \times u^T \right)
\times (\theta \times \Omega_L)^T \\
& = (\theta \times \Omega_L ) \times (\theta \times \Omega_L)^T 
\text{, as }
u_k \sim N(0,1)\\
& = \theta \times (\Omega_L \times \Omega_L^T) \times \theta\\
& \myeq \theta \times \Omega \times \theta\\
& \myeq \Sigma.
\end{align*}

A last point that can be optimized is the sampling from the Cauchy distribution
\[
\theta_k \sim Cauchy(\lambda, \omega) \text{,\ \ \ constrained by } \theta_k > 0.
\]
Indeed, as mentioned by the Stan Development Team (2021, Stan Users Guide, chapters 1.13 and 23.7)\cite{stanRef2021}, sampling from heavy tailed distributions like the Cauchy distributions can be difficult with HMC.
The problem is linked to the fact that the tail requires a larger step size than the trunk when there is a heavy tail. If we take a small step size, the No-U-Turn sampler requires many steps when starting in the tail of the distribution, and if we take a large step size there will be too much rejection in the central zone of the distribution.
To mitigate this problem, the Stan Development Team proposes to do a reparametrization, like we did in Section \ref{firstreparam}.
The idea in this case is to sample from a uniformly distributed variable $W \sim uniform(0,1)$ and then use the inverse of the Cauchy cumulative distribution function to transform this variable:
\[
F^{-1}_{\theta_k}(W) = \lambda + \omega \cdot tan\left(\pi \left(  W - \frac{1}{2}\right) \right)
\]
As $W \sim uniform(0,1)$, we have that $F^{-1}_{\theta_k}(W) \sim Cauchy(\lambda, \omega)$.

\newpage
\section{Implementation and results of the Bayesian hierarchical analysis}
\label{section4}

We will now present the results we obtained with our Bayesian hierarchical analysis.
To begin, we justify our choices for the implementation of the two Bayesian hierarchical models in the case of the multifaceted program.
Then, we present some sampling diagnostics of our Bayesian analysis.
After these first steps, we present the posterior distributions and the pooling we obtained for the parameters of interest.

\subsection{Application of our models to the multifaceted program}

The original analysis of the multifaceted program by Banerjee et al. (2015)\cite{banerjeeC} contains measures of several outcomes.
We decided to restrict our analysis to the outcome of total asset ownership 24 months after the start of the intervention.
This choice is motivated by two different reasons.
First, as mentioned in Section \ref{openQ}, Banerjee et al. observed an heterogeneity of the results between the different sites for this outcome, and it would therefore be interesting to see what knowledge our hierarchical models bring us about this heterogeneity.
The second reason is more fundamental. As the aim of the multifaceted program is for its authors to extract people from extreme poverty by the combination of an asset transfer with others interventions, it is important to understand whether the asset transfer substantially and durably increases the value of total asset ownership of an household.

As explained in the supplementary materials by Banerjee et al. (2015)\cite{banerjeeCSupp}, our outcome of interest, named by the authors total asset index, represents the total value of all durable assets and the value of all livestock assets owned by an household combined.
This index is constructed using the value of goats as a unit.
For example, a bicycle in Ethiopia is expressed as 1.34 goats.
More details about the construction of the total asset index in each site are available in the supplementary materials by Banerjee et al. (2015)\cite{banerjeeCSupp} for interested readers.

In our analysis of the heterogeneity of the impact of the multifaceted program on household asset ownership, we use 3 different Bayesian hierarchical models.
Our first model, called \textbf{Model 1}, corresponds simply to the hierarchical model presented in Section \ref{FirstMod}, that uses the results of site-level regressions as input.
Our two other models, called \textbf{Model 2} and \textbf{Model 2bis}, follow instead the hierarchical model presented in Section \ref{secondmodelall}, that uses the full dataset.
The only difference between these two models is that Model 2 uses as site-level predictors the vector $X_{i} = (1, T_{i})$, with $T_{i}$ the ITT dummy for household $i$, while Model 2bis uses  $X_{i} = (1, T_{i}, y_{baseline,i})$, with $y_{baseline,i}$ being the baseline value of total asset index for household $i$.

For Models 2 and 2bis, we have to do an additional choice, about site-level predictors. 
We decide to use for both models the following site-level predictors $Z$: the value of the asset transfer (measured in local goat price) and the presence of a health component in the program.
We did not introduce other site-level predictors, as we found that other site-level information available was difficult to convert into a comparable predictor for all sites.

\subsection{Sampling diagnostics for HMC}

Without implementing the optimization methods presented in Sections \ref{firstreparam} and \ref{optideux}, the Stan sampling diagnostics indicate that there is an important proportion of divergent transitions during the HMC process. This is problematic as it is an indicator that the returned posterior estimates can be biased.
According to the Stan Development Team (2020)\cite{stanRef2021}, and Gabry and Modrák (2021)\cite{Gabry2021}, the presence of divergent transitions during the exploration by HMC of the target posterior distribution might be due to the use of a too big step size for the exploration of the possibly small features of the target distribution: they therefore recommend to use a smaller step size, which we try with our data.

As the issue does not disappear even with a smaller step size, we have to address it with the optimization methods presented in Sections \ref{firstreparam} and \ref{optideux}.
This operation is successful, as the results obtained with the optimized models no longer indicate divergent transitions.
\\In order to monitor whether our HMC chains converge to the equilibrium distribution with our optimized models, we use the potential scale reduction $\hat{R}$, a statistic presented by the Stan Development Team (2020)\cite{stanRef2021} and Gelman et al. (2013, page 284)\cite{gelman2013} that is equal to 1 in the case where convergence is reached, and superior to 1 otherwise.
We obtain values equal to 1 for our different models, indicating therefore no convergence issues in our HMC sampling.
\\To conclude, Stan sampling diagnostics do not indicate any sampling issue with our optimized models.

\subsection{Posterior distributions obtained with the different models}
\label{PosteriorChap}

\begin{figure}[h!]
\caption{Density of posterior distributions of the ITT effect of the multifaceted program on total asset ownership obtained with Models 1, 2 and 2bis. The blue area corresponds to the values included into the 95\% credible intervals.}
\label{Fig::GrandeFigure}
\centering
\includegraphics[width=.49\textwidth]{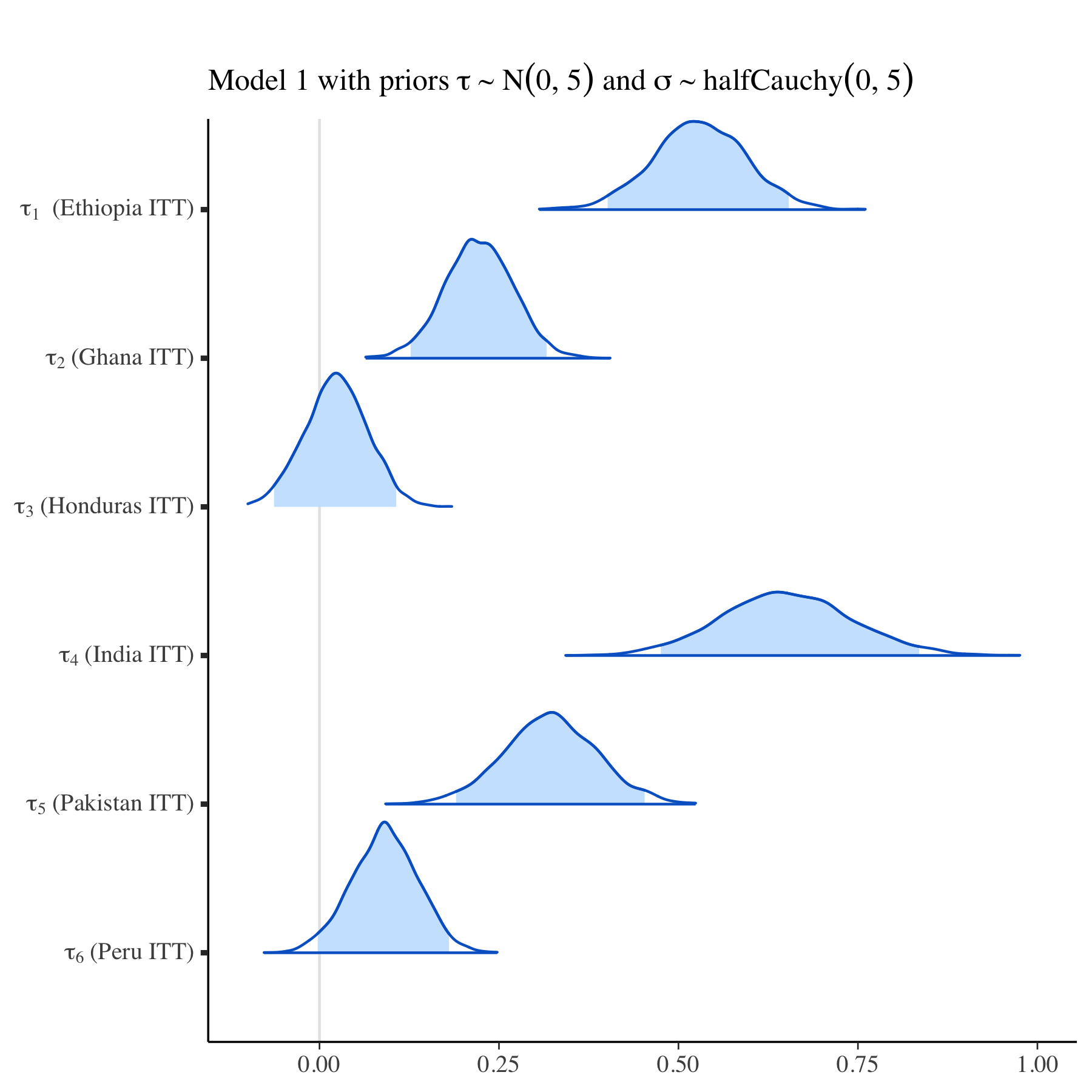}
\includegraphics[width=.49\textwidth]{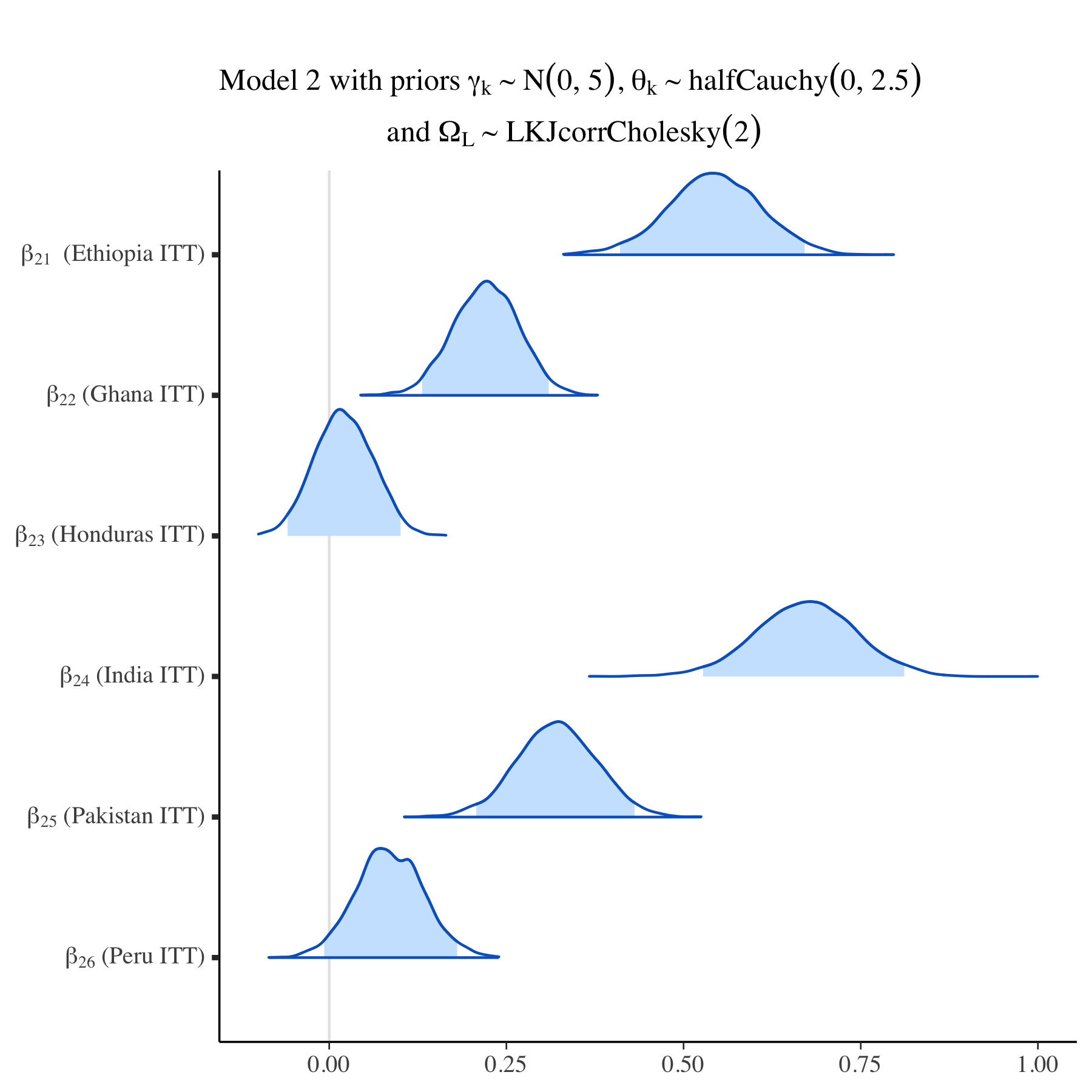}
\vspace{0.1cm}
\includegraphics[width=.55\textwidth]{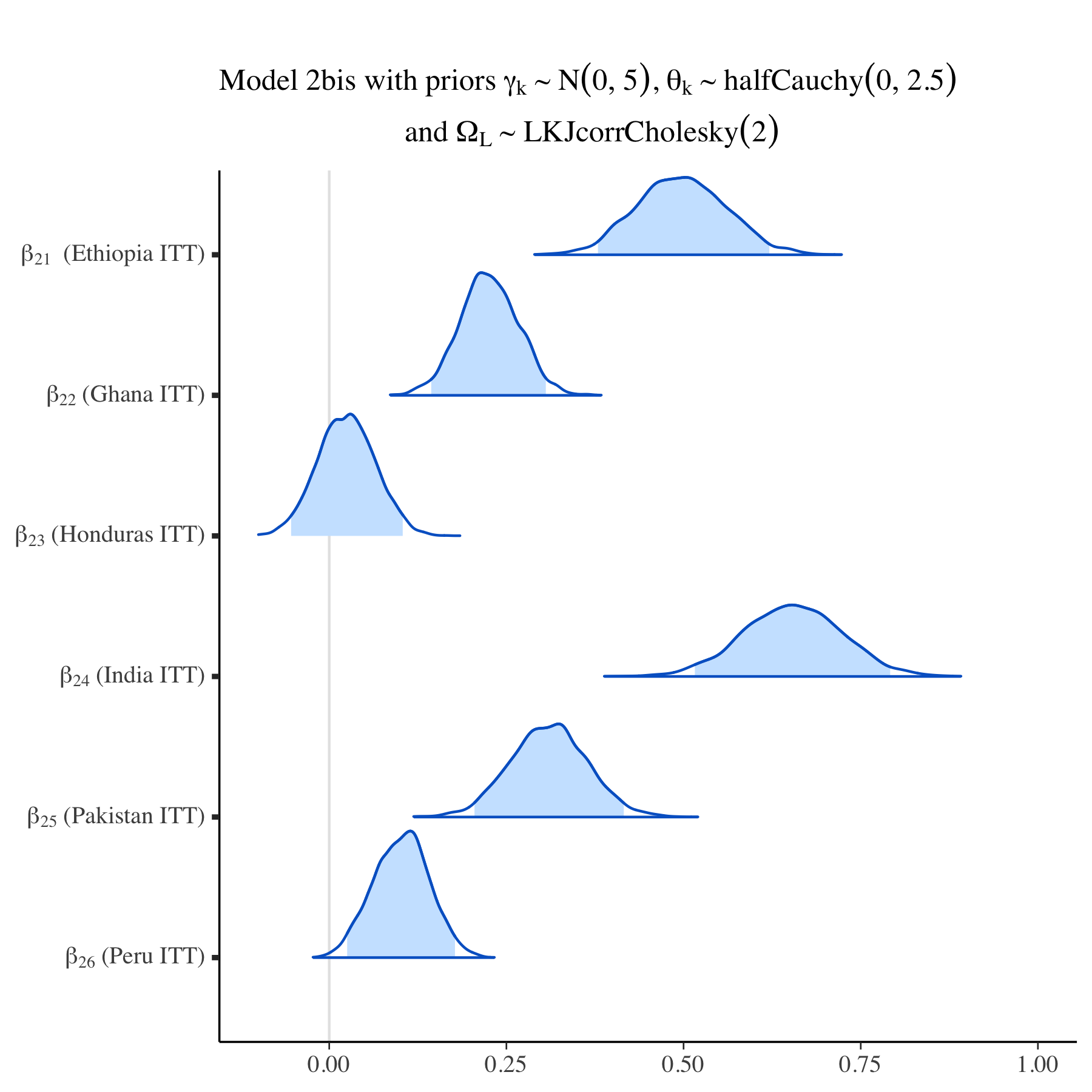}
\end{figure}

We present the posterior distributions obtained for the ITT effects of the multifaceted program on household total asset ownership for our 3 models  graphically in Figure \ref{Fig::GrandeFigure} and numerically in the Tables \ref{Table::Model1}, \ref{Table::Model2} and \ref{Table::Model2bis}.
We also present the estimates and standard errors for these ITT effects obtained with the separate site-level regressions (No-Pooling) in Table \ref{Table::NoPooling}.

\begin{table}[h!]
\caption{Posterior distributions of the ITT effect of the multifaceted program on total asset ownership obtained with Model 1 for each site}
\label{Table::Model1}
\centering
\begin{tabular}{lrrrrrrr}
  \toprule
Parameter & Mean & Standard deviation & 2.5\% quantile & 25\% & 50\% & 75\% & 97.5\% \\
  \midrule
  $\tau_1$ (Ethiopia) & 0.53 & 0.06 & 0.40 & 0.49 & 0.53 & 0.58 & 0.66 \\ 
  $\tau_2$ (Ghana) & 0.22 & 0.05 & 0.13 & 0.19 & 0.22 & 0.25 & 0.32 \\ 
  $\tau_3$ (Honduras) & 0.02 & 0.04 & -0.06 & -0.01 & 0.02 & 0.05 & 0.11 \\ 
  $\tau_4$ (India) & 0.65 & 0.09 & 0.48 & 0.59 & 0.65 & 0.71 & 0.83 \\ 
  $\tau_5$ (Pakistan) & 0.32 & 0.06 & 0.20 & 0.28 & 0.32 & 0.36 & 0.45 \\ 
  $\tau_6$ (Peru) & 0.09 & 0.05 & 0.00 & 0.06 & 0.09 & 0.12 & 0.18 \\ 
  $\tau$ (Hyper-parameter) & 0.31 & 0.15 & -0.01 & 0.22 & 0.31 & 0.39 & 0.62 \\ 
  \bottomrule
\end{tabular}
\end{table}

\begin{table}[h!]
\caption{Posterior distributions of the ITT effect of the multifaceted program on total asset ownership obtained with Model 2 for each site}
\label{Table::Model2}
\centering
\begin{tabular}{lrrrrrrr}
  \toprule
Parameter & Mean & Standard deviation & 2.5\% quantile & 25\% & 50\% & 75\% & 97.5\% \\ 
  \midrule
  $\beta_{2,1}=\tau_1$ (Ethiopia) & 0.54 & 0.07 & 0.41 & 0.50 & 0.54 & 0.59 & 0.67 \\ 
  $\beta_{2,2}=\tau_2$ (Ghana) & 0.22 & 0.05 & 0.13 & 0.19 & 0.22 & 0.25 & 0.31 \\ 
  $\beta_{2,3}=\tau_3$ (Honduras) & 0.02 & 0.04 & -0.06 & -0.01 & 0.02 & 0.05 & 0.10 \\
  $\beta_{2,4}=\tau_4$ (India) & 0.67 & 0.07 & 0.53 & 0.63 & 0.67 & 0.72 & 0.81 \\ 
  $\beta_{2,5}=\tau_5$ (Pakistan) & 0.32 & 0.06 & 0.21 & 0.28 & 0.32 & 0.36 & 0.43 \\ 
  $\beta_{2,6}=\tau_6$ (Peru) & 0.09 & 0.05 & -0.01 & 0.05 & 0.08 & 0.12 & 0.18 \\ 
  \bottomrule
\end{tabular}
\end{table}

\begin{table}[h!]
\caption{Posterior distributions of the ITT effect of the multifaceted program on total asset ownership obtained with Model 2bis for each site}
\label{Table::Model2bis}
\centering
\begin{tabular}{lrrrrrrr}
  \toprule
Parameter & Mean & Standard deviation & 2.5\% quantile & 25\% & 50\% & 75\% & 97.5\% \\ 
  \midrule
  $\beta_{2,1}=\tau_1$ (Ethiopia) & 0.50 & 0.06 & 0.38 & 0.45 & 0.50 & 0.54 & 0.62 \\ 
  $\beta_{2,2}=\tau_2$ (Ghana) & 0.23 & 0.04 & 0.14 & 0.20 & 0.22 & 0.25 & 0.31 \\ 
  $\beta_{2,3}=\tau_3$ (Honduras) & 0.02 & 0.04 & -0.05 & -0.00 & 0.02 & 0.05 & 0.10 \\
  $\beta_{2,4}=\tau_4$ (India) & 0.65 & 0.07 & 0.52 & 0.61 & 0.65 & 0.70 & 0.79 \\ 
  $\beta_{2,5}=\tau_5$ (Pakistan) & 0.31 & 0.05 & 0.20 & 0.27 & 0.31 & 0.35 & 0.42 \\ 
  $\beta_{2,6}=\tau_6$ (Peru) & 0.10 & 0.04 & 0.03 & 0.07 & 0.10 & 0.13 & 0.18 \\ 
   \bottomrule
\end{tabular}
\end{table}

\begin{table}[h!]
\caption{Estimates and standard errors for the ITT effect of the multifaceted program on total asset ownership obtained with site-level regressions (No-Pooling)}
\label{Table::NoPooling}
\centering
\begin{tabular}{lrr}
  \toprule
Parameter & Estimate & Standard error \\
  \midrule
  $\tau_1$ (Ethiopia) & 0.54 & 0.07  \\ 
  $\tau_2$ (Ghana) & 0.22 & 0.05  \\ 
  $\tau_3$ (Honduras) & 0.02 & 0.04  \\ 
  $\tau_4$ (India) & 0.69 & 0.09  \\ 
  $\tau_5$ (Pakistan) & 0.32 & 0.07  \\ 
  $\tau_6$ (Peru) & 0.08 & 0.05  \\ 
  \bottomrule
\end{tabular}
\end{table}


Let us first note that direct inference from the posterior distributions obtained with our models is possible, as we are in a Bayesian framework.
As highlighted by Sorensen et al. (2016)\cite{sorensen2016}, this is one of the important advantages of Bayesian statistics compared to frequentist statistics.
For instance, we can tell from Table \ref{Table::Model1} that approximately 95\% of the posterior density $\tau_3$ lies between the 2.5th percentile -0.06 and the 97.5th percentile 0.11, according to Model 1.
We can also tell that there is approximately probability of 25\% that $\tau_3$ is inferior to -0.01 for this model.
It can be seen in Table \ref{Table::Model2} that this probability is the same according to Model 2.

After this first remark, we can also highlight that, for all the sites, the means and quantiles of the posterior distributions of the ITT effects $\tau_s$ are very similar between Model 1 and more complex Model 2.
However, there are some differences for Model 2bis, that remain very small: for instance, the differences with Model 2 of the mean of the posterior distributions for the different $\tau_s$ are always smaller than the standard deviation of the posterior distributions.
These small differences of Model 2bis can probably be explained by the introduction in this model of baseline values of household asset ownership, that are not present in Model 2 nor in the construction of site-level estimates and standard errors of Table \ref{Table::NoPooling} that are used as input in Model 1.

Another interesting observation is that the means of the posterior distributions of the ITT effects $\tau_s$ obtained with Models 1 and 2 are very close from the estimates for $\tau_s$ obtained with the simple site-level OLS regressions, that is with the no-pooling model.
Therefore, at first glance, our Bayesian hierarchical analysis brings very close results about the impact of the multifaceted program on asset ownership 24 months after the beginning of the intervention compared to the original no-pooling analysis by Banerjee et al. (2015)\cite{banerjeeC}.
Indeed, similarly to the original study, we find with all the models a significantly positive impact of the intervention on asset ownership, unless for Honduras where the value of 0 is included into the 95\% credible intervals.

We will come back in Section \ref{section5} on the interpretation of the similarity of the results of our analysis compared to the original analysis, but let us before quantify the level of pooling obtained with our hierarchical analysis.

\subsection{Pooling with our different hierarchical models}
\label{poolingChap}

Given the important proximity of the results for Models 1 and 2, we will focus on the analysis of information pooling with Model 1 only.
To evaluate the pooling of information between the different sites, we use the approach proposed by Gelman and Pardoe (2006)\cite{gelmanPardoe2006}.
This approach, also used by Meager (2019)\cite{meager}, will enable us to compare our results with the results she found by applying a Bayesian hierarchical analysis to study the impact of microcredit expansion interventions.

In their very interesting paper, Gelman and Pardoe (2006)\cite{gelmanPardoe2006} propose what they call a \textbf{pooling factor}, that represents for each site $s$ the extent of information pooling with other sites. 
In the case of Model 1, it can be written as:
\[
\omega_s = \frac{\Hat{\sigma}_s^2}{\Tilde{\sigma}^2 + \Hat{\sigma}_s^2},
\]
with $\Tilde{\sigma}$ the mean of the posterior distribution of $\sigma$ obtained with our hierarchical model, and $\Hat{\sigma}_s^2$ the sampling error defined in Section \ref{FirstMod}.

To have an intuition of what this pooling factor $\omega_s$ represents, let us
take two extreme cases.
In the case where our posterior results indicate no heterogeneity of the ITT effect $\tau_s$ between sites, we have that $\Tilde{\sigma}^2$ is very close to 0. This leads to a pooling factor very close to $\omega_s = 1$, meaning that we have a full-pooling of information between sites.
This is quite intuitive: as our posterior results indicate no heterogeneity of the ITT effect $\tau_s$ between sites, the effect we measure in India should tell us exactly what happens in another site like Bangladesh, so there is a lot of pooling.
On the contrary, in the case where our posterior results indicate a very important heterogeneity of the ITT effect $\tau_s$ between sites compared to the sampling error in site $s$, we have that $\Tilde{\sigma}^2 >> \Hat{\sigma}_s^2$.
This leads to a pooling factor close to $\omega_s = 0$, meaning that we have no pooling of information between sites.
This is again very intuitive: as our posterior indicates very much heterogeneity of the ITT effect $\tau_s$ between sites, the effect we measure in India does not provide us much information about what happens in another site like Bangladesh, so there is little pooling.

\begin{figure}[h!]
\caption{Density of posterior distributions of the between sites heterogeneity parameter $\sigma$ obtained with Model 1. The blue area corresponds to the values included into the 95\% credible intervals.}
\label{Fig::FigureSigma}
\centering
\vspace{-0.8cm}
\includegraphics[width=.60\textwidth]{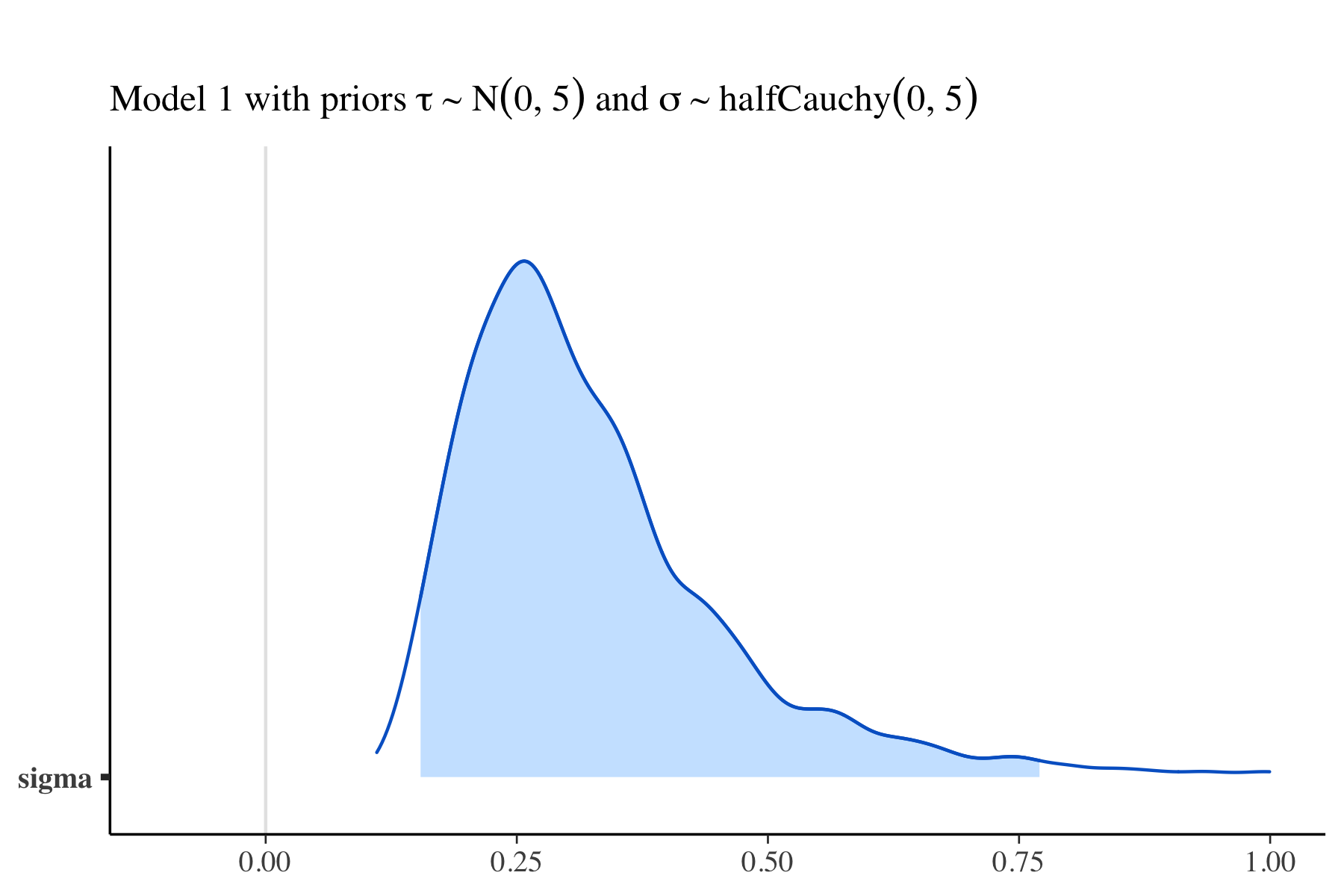}
\end{figure}

After this quick introduction of the pooling factor $\omega_s$, we present the the posterior distributions obtained for the between sites heterogeneity parameter $\sigma$ of our Model 1 graphically in Figure \ref{Fig::FigureSigma} and numerically in Table \ref{Table::Model1Sigma}.
We also present the pooling factors $\omega_s$ obtained with this model in each site $s$ in Table \ref{Table::Model1Pooling}.

\begin{table}[h!]
\caption{Posterior distributions of the heterogeneity between sites parameter $\sigma$ obtained with Model 1}
\label{Table::Model1Sigma}
\centering
\begin{tabular}{lrrrrrrr}
  \toprule
Parameter & Mean & Standard deviation & 2.5\% quantile & 25\% & 50\% & 75\% & 97.5\% \\ 
  \midrule
  $\sigma$ & $\Tilde{\sigma} = 0.34$ & 0.18 & 0.15 & 0.23 & 0.30 & 0.39 & 0.77 \\ 
   \bottomrule
\end{tabular}
\end{table}

\begin{table}[h!]
\caption{Pooling parameter $\omega_s$ and reminder of input $\hat{\sigma}_s$ for each site $s$ obtained with Model 1}
\label{Table::Model1Pooling}
\centering
\begin{tabular}{lrrr}
  \toprule
Site & $\hat{\sigma}_s$ & $\omega_s$  \\ 
  \midrule
  Ethiopia & 0.066 & 0.04  \\ 
  Ghana & 0.048 & 0.02  \\ 
  Honduras & 0.044 & 0.02  \\ 
  India & 0.090 & 0.07  \\ 
  Pakistan & 0.067 & 0.04  \\ 
  Peru & 0.047 & 0.02  \\ 
   \bottomrule
\end{tabular}
\end{table}

We can notice from these results that our hierarchical Model 1 leads to pooling factors $\omega_s$ ranging from 2\% to 7\% in the different sites.
This is a very small amount of pooling compared to the ones found by Meager (2019)\cite{meager} during her analysis of microcredit expansion, which were around 50 \%, with variations according to the outcomes observed.

In the next Section, we will try to understand where these differences of pooling between our results with the multifaceted program and Meager's results with microcredit expansion come from.
We will also try to understand why Model 1 and Model 2 give very similar results.

\newpage
\section{Interpretation of the results}
\label{section5}

We will now provide an interpretation of the main results we obtained with our Bayesian hierarchical analysis.
To begin, we analyze the important differences of our results with the ones of the first application of Bayesian hierarchical analysis in development economics by Meager (2019)\cite{meager}.
Then, we focus on the interpretation of the important similarity of the results we obtained with Model 1 and more complex Model 2.
Finally, we conclude on the knowledge that this Bayesian hierarchical analysis brings us about the multifaceted program.

\subsection{Difference with the results obtained by Meager (2019)}

In Section \ref{poolingChap}, we have noticed that our hierarchical Model 1 leads to pooling factors $\omega_s$ ranging from 2\% to 7\% in the different sites, which correspond to an average on the different sites of $\Bar{\omega} = 3\%$.
This average pooling $\Bar{\omega}$ is much lower than the ones obtained by Meager (2019)\cite{meager} in her study of microcredit expansion, ranging from 30\% to 50\% depending on the variable observed.

To understand if this difference of our results with Meager's ones comes from the model we use or more fundamentally from the data we analyze, we apply our Bayesian hierarchical Model 1 to her microcredit data, and find pooling averages $\Bar{\omega}$ of the same order of magnitude as those she found.
Therefore, the difference with the results of the Bayesian hierarchical analysis by Meager does not seem to come from the model, but rather from the data.

Compared to the site-level regressions of the multifaceted program, the ones of the microcredit expansion program lead to treatment effects for which the standard errors $\Hat{\sigma}_s$ are of the same magnitude than the estimates $\Hat{\tau}_s$.
This is linked to the fact that the effects measured for the microcredit expansion program are closer to 0 (and less significant) than the effects measured for the multifaceted program.
To understand better how the amount of pooling changes with the properties of our data, we decide to run some simulations.

As the data input in our Model 1, excluding the weakly informative priors, consists only of the estimates of the treatment effect $\Hat{\tau}_s$ and their associated standard errors $\Hat{\sigma}_s$ obtained with site-level regressions, we decide to run some simulations by increasing and decreasing $\Hat{\tau}_s$ and $\Hat{\sigma}_s$ respectively, and see what happens to the average amount of pooling.
The results of these simulations are presented in Table \ref{Table::Model1Simu}.

\begin{table}[h!]
\caption{Pooling parameter average $\Bar{\omega}$ over all sites and posterior mean $\Tilde{\sigma}$ of parameter $\sigma$ obtained with Model 1 for different values of $\Hat{\tau}_s$ and $\Hat{\sigma}_s$ ("Original", "Original $\times$ 10", and "Original $\div$ 10" indicating respectively the original value obtained with the multifaceted program data, this value multiplied by ten, and divided by ten)}
\label{Table::Model1Simu}
\centering
\begin{tabular}{lrrr}
  \toprule
Value of $\Hat{\tau}_s$ & Value of $\Hat{\sigma}_s$  & $\Tilde{\sigma}$ & $\Bar{\omega}$  \\ 
  \midrule
  Original & Original & 0.34 & 0.03  \\ 
  Original $\times$ 10 & Original & 3.34 & 0.0003  \\ 
  Original $\div$ 10 & Original & 0.03 & 0.78  \\ 
  Original & Original $\times$ 10 & 0.31 & 0.77  \\ 
  Original & Original $\div$ 10 & 0.38 & 0.0003  \\
  All $\Hat{\tau}_s$ equal to original $\Hat{\tau}_{Ethiopia}$ & Original & 0.31 & 0.7681159  \\
   \bottomrule
\end{tabular}
\end{table}

It can be seen in Table \ref{Table::Model1Simu} that our simulations lead to a bigger amount of pooling when we increase the value of $\Hat{\sigma}_s$ with respect to the value of $\Hat{\tau}_s$.
This result corresponds well to what we observed for Meager's results: microcredit data leads to bigger values of $\Hat{\sigma}_s$ with respect to $\Hat{\tau}_s$ in site-level regressions than the multifaceted program, and this leads indeed to a bigger amount of pooling.
We can also notice from Table \ref{Table::Model1Simu} that we have an important pooling for our simulation with $\Hat{\tau}_s$ values all equal to the original one of Ethiopia.

So, according to our simulations, the first order behavior of our Bayesian hierarchical Model 1 seems to be the following.
When the the site-level estimates $\Hat{\tau}_s$ are close enough to one another comparatively to their associated standard errors, the model seems to consider that the observed heterogeneity is mostly due to sampling error of the ITT effect measure in each site and will therefore enact a high amount of pooling of information between sites.
On the contrary, when the site-level estimates $\Hat{\tau}_s$ are far from one another (always using as metric their associated standard errors), the model seems to consider that the observed heterogeneity is mostly due to true heterogeneity of the ITT effect between sites and will not enact a high amount of pooling of information between sites.

\subsection{Similarity of the results obtained with the different models}

In Section \ref{PosteriorChap}, we have noticed that our hierarchical models 1 and 2 gave very similar posterior distributions of the ITT effects $\tau_s$.
Given the fact that Model 2 uses more information than Model 1, with the presence of site-level predictors and household-level data, one could have thought that the two models would lead to different results.
The important similarity of the results obtained with these models is maybe linked to the fact that there is very low pooling of information in the case of the multifaceted program, as detailed in previous section.
Indeed, in both models, the posterior means of $\tau_s$ are extremely close to the initial site-level predictors $\Hat{\tau}_s$.
Therefore, this similarity of the results obtained with simpler Model 1 and more complex Model 2 should not be taken as a general result: the two models could lead to different results with other datasets that lead to stronger pooling.
Thus, for the study of other programs, it can always be interesting to also implement the more detailed Model 2 when the required data is available.

\subsection{What did this analysis bring to our knowledge about the multifaceted program?}
 
To conclude, our Bayesian hierarchical analysis leads to a very low amount of pooling of information between sites.
This gives us estimates of the ITT effect of the multifaceted program on asset ownership that are very close to the ones of the simple no-pooling site-level regressions.
According to our different models, leading all to similar results, the observed heterogeneity of the site-level estimates of the impact of the multifaceted program on asset ownership 24 months after the asset transfer should be interpreted mainly as "true" inter-site heterogeneity. 
Indeed, according to the pooling factor calculated for Model 1, our posterior estimate $\Tilde{\sigma}$ of $\sigma$ is much bigger than the sampling variation $\hat{\sigma}_s$.
This means that $\hat{\tau}_s$ is a better signal of $\tau_s$ than $\tau_s$ is of $\tau$. 
More concretely, what we learn in the sites different from Honduras, where the asset ownership increased significantly 24 months after the asset transfer, does not impact much our posterior belief about what happens in Honduras, where the effect measured was close to zero. 
Thus, according to our models, the disappointing results in Honduras reflect true heterogeneity between sites, and are not simply due to sampling error.

In the next Section, we will see some ideas of further research that could be used to deepen the analysis of the multifaceted program.

\newpage
\section{Ideas of further research}
\label{section6}
We will finally give some ideas of further research about the multifaceted program, that could be explored in future work.
First, we discuss about the possibility to include more informative priors.
Then, we mention some further methods relative to model comparison.
Finally, we propose some ideas and their related challenges about the modelling of the interaction between the different outcomes of the multifaceted program, and about the inclusion of more complexity in our models.

\subsection{Inclusion of informative priors based on similar studies}

In Section \ref{autrePourPrior}, we have mentioned that a program similar to the multifaceted program of Banerjee et al. (2015)\cite{banerjeeC} has been studied by Bandiera et al. (2017)\cite{bandiera2017}.
In our analysis, we decided not to include information of previous studies.
Let us note that we could have done differently if we had the belief that the results of the analysis of Bandiera et al. (2017)\cite{bandiera2017} could bring some information to our analysis. 

More precisely, for our Model 1, we have different possible prior choices for $\tau$ and $\sigma$, that depend on our beliefs.
In our analysis, we did not want to include information of previous studies, and therefore assigned a weakly informative prior on $\tau$ and $\sigma$.
However, it could have been done differently: one could believe that the analysis by Bandiera et al. (2017)\cite{bandiera2017} provides enough information to set more informative priors on $\tau$ and $\sigma$.
It would be interesting to study in a future work how and when to include information of other studies into the prior, and see how much it changes the results of the analysis.

\subsection{Model selection by predictive performance evaluation}

In our analysis, we obtained very similar results with our different models.
However, it might happen that these models lead to diverse results for other outcomes of the multifaceted program or even for the analysis of another program.
In this case, we would be further interested by the selection of one model between our different proposals, and we should use model selection methods.

As highlighted by Vehtari et al. (2016)\cite{vehtari2016} and Haugh (2021)\cite{haugh2021}, several model selection methods exist. 
They are all based on the estimation of the out-of-sample predictive accuracy of the different models, and select the one with the best predictive accuracy.
More details about the different model selection methods and their implementation are available in the very interesting paper by Vehtari et al. (2016)\cite{vehtari2016} for interested readers.
Let us note that Vehtari et al.(2020)\cite{vehtari2020} developed an R package that permits to implement these model selection methods on Stan functions outcomes.

\subsection{Modeling the interaction of the different outcomes in the multifaceted program}

In our Bayesian hierarchical analysis, we have focused on the impact of the multifaceted program on one outcome only: the total asset ownership.
If we have justified this choice for our analysis, it would be interesting, in future work about the multifaceted program, to study the interaction of the different outcomes and try to understand how they interact together. 
It would permit to further understand the mechanisms of the multifaceted program.
For instance, we could try to model the impact of the multifaceted program on the different outcomes through a hierarchical model that divides the outcomes into different families, representing the different channels through which the program can fail.
We could then evaluate our different model propositions with the model selection methods presented before.

Let us note that such an analysis presents however an important challenge.
Such modelling would lead to complex models, with more levels than the models we have implemented in our analysis with one outcome only.
Indeed, there would be not only an hierarchy due to the presence of different sites, but also an hierarchy due to the presence of different outcomes.
Given that our two-level models, that were much more simpler, already required optimization methods to obtain a good sampling with our HMC, we can imagine that more complex models would face even more computational issues, that could maybe prevent to carry out the analysis.

\subsection{Further inclusion of complexity}
\label{moreComplex}

As mentioned in Section \ref{FirstMod}, we departed in our Model 1 from simple site-level regressions of the outcome of interest $Y_i$ on the ITT indicator $T_i$.
These site-level regressions led to estimates of the ITT effects $\Hat{\tau}_s$
and associated standard errors $\Hat{\sigma}_s$ very close from the ones presented in the no-pooling analysis by Banerjee et al. (2015)\cite{banerjeeC}.
Let us highlight however that the original paper introduced in its site-level regressions additional covariates, such as baseline values of the outcome of interest and other household variables.

We decided to depart from our simpler regression for different reasons.
First, it would have been computationally very difficult to include all the covariates used by the authors in Model 2, in order to carry out a comparison between Model 1 and 2. 
Indeed, we experimented that adding several household-level covariates in Model 2 leads to divergences issues in the HMC process, and thus to a bad quality sampling preventing the hierarchical analysis.
Second, our estimates of $\Hat{\tau}_s$ and $\Hat{\sigma}_s$ are extremely close to the ones of the original study presented in the supplementary materials by Banerjee et al. (2015)\cite{banerjeeCSupp}: we checked with Model 1 that this very small difference does not impact the results of our analysis.
However, it would be interesting in future work to address the challenge of the inclusion of further complexity in the models we presented.

\newpage
\section{Conclusion}

To conclude, the Bayesian hierarchical analysis we led with our two different models brings a very low amount of pooling of information between sites.
This gives us very close results about the impact of the multifaceted program on asset ownership 24 months after the beginning of the intervention compared to the original no-pooling analysis by Banerjee et al. (2015)\cite{banerjeeC}.
Indeed, similarly to the original study, we find with all the models a significantly positive impact of the intervention on asset ownership for all sites except Honduras, where the value of 0 is included into the 95\% credible intervals.

According to our different models, leading all to similar results, the observed heterogeneity of the site-level estimates of the impact of the multifaceted program on asset ownership 24 months after the asset transfer should be interpreted mainly as true inter-site heterogeneity. 
Indeed, according to the pooling factor calculated for Model 1, our posterior estimate $\Tilde{\sigma}$ of $\sigma$ is much bigger than the sampling variation $\hat{\sigma}_s$.
This means that $\hat{\tau}_s$ is a better signal of $\tau_s$ than $\tau_s$ is of $\tau$. 
More concretely, what we learn in the sites different from Honduras, where the asset ownership increased significantly 24 months after the asset transfer, does not impact much our posterior belief about what happens in Honduras, where the effect measured was close to zero. 
Thus, according to our models, the disappointing results in Honduras reflect true heterogeneity between sites, and are not simply due to sampling error.

These results suggest that a very big part of the observed heterogeneity of the impact between sites is due to true heterogeneity between sites.
This differs importantly from the results of the first application of Bayesian hierarchical analysis in development economics by Meager (2019)\cite{meager}.
With her study about microcredit expansion, she found levels of pooling between sites way more important than the ones we found, suggesting that an important proportion of the observed heterogeneity was due to sampling error in her case.

The results of our simulations permit us to understand the reason of this difference.
In fact, the Bayesian hierarchical models we used tend to have the following first order behavior.
When the the site-level estimates of the program impact are close enough to one another comparatively to their associated standard errors, the model seems to consider that the observed heterogeneity is mostly due to sampling error of the ITT effect measure in each site, and will therefore enact a high amount of pooling of information between sites.
This is the case with Meager's analysis.
On the contrary, when the site-level estimates of the program impact are far from one another (always using as metric their associated standard errors), the model seems to consider that the observed heterogeneity is mostly due to true heterogeneity of the ITT effect between sites, and will not enact a high amount of pooling of information between sites.
This is the case with our analysis.

Let us finally note that the important similarity of the results obtained with our models 1 and 2 is maybe linked to the fact that there is very low pooling of information in our case.
Therefore, this similarity of the results obtained with simpler Model 1 and more complex Model 2 should not be taken as a general result: the two models could lead to different results with other datasets that lead to stronger pooling.
Thus, for the study of other programs, it can always be interesting to also implement the more detailed Model 2 when the required data is available.

\phantomsection 
\addcontentsline{toc}{section}{References} 
\newgeometry{left=3cm} 
\printbibliography
\restoregeometry 

\newpage
\section{Appendices}

\subsection{Appendix 1: Methods for the implementation on R and Stan}
\appendices{Appendix 1: Methods for the implementation on R and Stan}

The full code is available on GitHub at \url{https://github.com/louischarlot/Bayesian_hierarchical_analysis_multifaceted_program_extreme_poverty}.

We want to provide here some practical information about the implementation of our Bayesian hierarchical analysis on R and Stan.

The first step is to download RStan, the R version of Stan, following the steps indicated at \url{https://github.com/stan-dev/rstan/wiki/RStan-Getting-Started}.
Let us note for interested readers that Stata and Python versions of Stan are also available.

Once the installation is completed, we have to create different Stan files to write the hierarchical models we want to sample with HMC for our Bayesian analysis.
Let us note that a Stan file will be written in the C++ programming language, as it will be a C++ compiler that will carry out the HMC sampling, communicating with R to import the input data and to export the results.

The code for Model 1 can be included in a Stan file \textit{Model\_1.stan}. Using the same variable notations as in Section \ref{FirstMod}, it can be written as follows:
\lstset { 
    language=C++,
    backgroundcolor=\color{black!5}, 
    basicstyle=\footnotesize,
    commentstyle=\color{teal}
}

\begin{customFrame}
//  The data block specifies the data that is conditioned upon in Bayes Rule.
data {       
  int<lower=0> S;  // number of sites: must be positive
  real hat_tau_s[S];  // vector of estimated treatment effects
  real<lower=0> sigma_s[S];  // vector of standard errors of estimated treatment effects: components must all be positive
}

// The parameters block declares the parameters which posterior distribution is sought.
parameters {         
  real tau;  // mean of the treatment effect (over all sites)
  real<lower=0> sigma;   // standard deviation of the treatment effect (over all sites)
  vector[S] eta;    // tau_s = mu + tau * eta is the mean of the treatment effect (for each site)
}

// We use a "non-centered parameterization":
transformed parameters { 
  vector[S] tau_s;
  tau_s = tau + sigma * eta;
}

// We finally write our hierarchical model, with the priors:
model {  
  tau ~ normal(0, sqrt(5)); // hyperprior 1 : N(0,5)
  sigma ~ cauchy(0, 5); // hyperprior 2 : half-Cauchy(0,5)
  target += normal_lpdf(eta | 0, 1);    
  target += normal_lpdf(hat_tau_s | tau_s, sigma_s);
}
\end{customFrame}

The code for Model 2 can be included in a Stan file \textit{Model\_2.stan}. Using the same variable notations as in Section \ref{secondmodelall}, it can be written as follows:

\begin{customFrame}
//  The data block specifies the data that is conditioned upon in Bayes Rule.
data {
  int<lower=0> N;    // number of  individuals
  int<lower=1> I;    // number of individual predictors
  int<lower=1> S;    // number of site
  int<lower=1> J;    // number of site predictors
  int<lower=1,upper=S> site[N];  // site of individual i
  matrix[N, I] X;    // individual-level predictors
  matrix[J, S] tZ;   // site-level predictors transposed
  vector[N] y;       // individual-level outcomes
}

// The parameters block declares the parameters which posterior distribution is sought.
parameters {
  matrix[I, S] u;
  cholesky_factor_corr[I] L_Omega;  // Cholesky factor for Omega
  vector<lower=0,upper=pi()/2>[I] theta_unif;  // uniform distribution used to build the Cauchy priors for theta
  matrix[I, J] gamma;   // site-level coefficients
  real<lower=0> sigma_s[S];  
}

// We use a "non-centered parameterization" for theta and
// a "Cholesky parametrization" for Omega:
transformed parameters {
  vector<lower=0>[I] theta = 2.5 * tan(theta_unif); //Cauchy priors for theta
  matrix[I, S] beta = gamma * tZ + diag_pre_multiply(theta, L_Omega) * u;
}

// We finally write our hierarchical model, with the priors:
model {
  vector[N] mu;
  vector[N] sigma; // variable used to merge the sigma_s of different sites
  sigma_s ~ uniform(0,100000);  // prior for sigma_s
  for(n in 1:N) {
    mu[n] = X[n, ] * beta[, site[n]];
    sigma[n] = sigma_s[site[n]];
  }
  to_vector(u) ~ std_normal(); 
  L_Omega ~ lkj_corr_cholesky(2);
  to_vector(gamma) ~ normal(0, sqrt(5)); // N(0,5)
  y ~ normal(mu, sigma);
}
\end{customFrame}

Once we have saved these models, we have to open an R file to run them on the data of the multifaceted program.
For example, in the case of Model 2, once we have downloaded our dataset from Banerjee et al. (2015)\cite{banerjeeC}, we can run our Bayesian hierarchical model in the R file \textit{Model\_2\_no\_baseline.R} as follows.

\lstset { %
    language=R,
    backgroundcolor=\color{black!5}, 
    basicstyle=\footnotesize,
    commentstyle=\color{teal}
}

We first prepare the data that will be used by the sampler:
\begin{customFrame}
y <- data_banerjee$asset_index_end # Vector of individual-level outcome of interest: the total asset ownership.
T <- data_banerjee$treatment # ITT indicator.
N <- length(y) # Number of individuals.
S <- 6  # Number of sites.
site <- data_banerjee$country # Site indicator. (1 = Ethiopia, 2 = Ghana, 3 = Honduras, 4 = India, 5 = Pakistan, 6 = Peru)
# We define the individual-level predictors:
data_banerjee$X_intercept <- rep(1,N)
Covariates_X <- dplyr::select(data_1, X_intercept, treatment, asset_index_bsl)
X <- as.matrix(Covariates_X) # Matrix of individual-level predictors.
I <- length(Covariates_X) # Number of individual-level predictors.
# We define the site-level predictors:
Z_intercept <- rep(1,S)
site_health_component <- c(0, 1, 1, 1, 1, 1)
site_value_asset_transfer <- c(7.98, 6, 4.75, 6.53, 3.75, 17.14) # in local goat price
Covariates_Z <- data.frame(Z_intercept, site_health_component, site_value_asset_transfer)
Z <- as.matrix(Covariates_Z) # Matrix of site-level predictors.
tZ <- t(Z) # We transpose the group predictors.
J <- length(Covariates_Z)  # Number of site-level predictors.

data_for_stan <- list(N = N, 
                      S = S, 
                      I = I,
                      J = J,
                      y = y,       
                      X = X,
                      Z = Z,
                      site = site)
\end{customFrame}

We then call the Stan file that contains our hierarchical model 2, to run the sampler on this prepared data:

\begin{customFrame}
fit_2 <- stan(
  file = "Model_2.stan", # Stan file that contains Model 2.
  data = data_for_stan,  # Data we have prepared for Stan.
  chains = 4,            # Number of Markov chains used by the sampler.
  warmup = 1000,         # Number of warmup iterations per chain.
  iter = 2000,           # Total number of iterations per chain.
  control=list(adapt_delta=0.99, max_treedepth = 14), # Reduce step size.
  cores = 4.             # Number of cores (we use one per chain if possible).
  
)
\end{customFrame}

We can finally observe the results of the Bayesian hierarchical analysis with the library bayesplot, presented by Gabry and Modrák (2021)\cite{Gabry2021}.

\subsection{Appendix 2: Data availability}
\appendices{Appendix 2: Data availability}

The data of the multifaceted program has been made available by the authors at \url{https://dataverse.harvard.edu/dataset.xhtml?persistentId=doi:10.7910/DVN/NHIXNT}.

We used in particular the household level data file \textit{pooled\_hh\_postanalysis.dta}, that can be obtained (in the folder ScienceDataRelease/data\_modified) by running the Stata do files of the authors (that are in the folder ScienceDataRelease/dofiles).

\subsection{Appendix 3: More details about MCMC and HMC}
\appendices{Appendix 3: More details about MCMC and HMC}

In this Appendix, we include some additional information about Markov Chain Monte Carlo (MCMC) and Hamiltonian Monte Carlo (HMC), inspired from the didactic explanations provided by Lambert (2018)\cite{Lambert}, Betancourt (2017)\cite{betancourt2017} and Neal (2011)\cite{neal2011}.

\subsubsection{Random-walk Metropolis-Hastings}
\label{MHalgo}
The first generation MCMC algorithm is the Metropolis-Hastings algorithm.
The idea of this algorithm is to obtain a sample from the posterior distribution
\[
\pi(\theta \mid y)
=\frac{p(y \mid \theta) \cdot \pi(\theta)} {\int_{\theta} p(y \mid \theta) \cdot \pi(\theta)}
\]
Given that the denominator of this expression is independent from $\theta$, the idea is to sample directly from $p(y \mid \theta) \cdot \pi(\theta)$.
As explained by Lambert (2018)\cite{Lambert}, this is done through the following steps:
\begin{itemize}
    \item First, we draw an initial value $\theta_0$ of $\theta$.
\end{itemize}
\begin{itemize}
    \item Then, for a large number $n$ of iterations, we repeat the following setps:
    \begin{itemize}
        \item \textbf{a)} We draw a proposal $\theta_{proposal}$ from a proposal distribution $q(\theta_{proposal} \mid \theta_{current\ step})$
        \item \textbf{b)} We calculate:
        $$\alpha 
        = \frac
        {p(y \mid \theta_{proposal}) \cdot \pi(\theta_{proposal})/ q(\theta_{proposal} \mid \theta_{current\ step})}
        {p(y \mid \theta_{current\ step}) \cdot \pi(\theta_{current\ step}))/ q(\theta_{current\ step} \mid \theta_{proposal})}$$
        We include $q(\theta_{current\ step} \mid \theta_{proposal})$ and $q(\theta_{proposal} \mid \theta_{current\ step})$ in order to correct the possible asymmetries of the proposal distribution $q$.
        \item \textbf{c)} Finally:
        \\If $\alpha > 1$, we accept the proposal ($\theta_{current\ step} = \theta_{proposal}$).
        \\If $0 < \alpha < 1$, we accept the proposal with probability $\alpha$, and reject it ($\theta_{current\ step} = \theta_{current\ step}$) with probability $1-\alpha$.
    \end{itemize}
\end{itemize}

A possible choice for the proposal distribution $q$ is a Normal distribution centered on $\theta_{current\ step}$: in this case, the algorithm is called the \textbf{Random-Walk Metropolis-Hastings}.
The advantage of using a Normal distribution is that it gives a simpler formula to calculate $\alpha$:
\[
\alpha 
= \frac
{p(y \mid \theta_{proposal}) \cdot \pi(\theta_{proposal})}
{p(y \mid \theta_{current\ step}) \cdot \pi(\theta_{current\ step})}
\]

\textbf{Random-Walk Metropolis-Hastings} has an important drawback: it is very \textbf{inefficient} when the posterior is multidimensional, like in our case.
Indeed, when the dimensionality of the posterior distribution increases, the new values $\theta_{proposal}$ proposed randomly around $\theta_{current\ step}$ by the algorithm at each step will often be located in zones of very low density of the posterior.
Therefore, a very low amount of these proposals will be accepted, leading to a very low sampling efficiency.
Therefore, we need a more efficient way to propose new values $\theta_{proposal}$: this is exactly the aim of HMC.

\subsubsection{Hamiltonian Monte Carlo (HMC)}
As highlighted by Lambert (2018)\cite{Lambert}, the difference of HMC with Metropolis-Hastings is in the \textbf{way we do proposals} $\theta_{proposal}$, that is \textbf{based on a physical analogy}.

The idea is to \textbf{attribute to a physical particle the position $\theta_{current\ step}$}, make it evolve in space for some time, and \textbf{take as proposal $\theta_{proposal}$ the new position of this particle in the space after this evolution}.

More precisely, we associate to a physical particle the energy:
\[E(\underbrace{\theta}_{position},\underbrace{M}_{momentum})
= \underbrace{U(\theta)}_{potential\ energy} + \underbrace{K(M)}_{kinetic\ energy}
= Constant
\]

The probability of the particle to be in the state $(\theta,M)$ is:
\[
p(\theta,M) \propto  e^{-(U(\theta)+K(M))/T}
\]
with:
\[
K(M) = \frac{1}{2} m \cdot v^2
= \frac{M^2}{2m}
\left(= \sum_{d=1}^{dim} \frac{M_d^2}{2m}\right)
\ \ and \ \ \
U(\theta)=-log\left( p(y \mid \theta) \cdot \pi(\theta) \right)
\]
Then, taking $m=1$ and $T=1$:
\[
p(\theta,M) \propto  p(y \mid \theta) \cdot \pi(\theta) \cdot 
\hspace{-0.5cm} 
\underbrace{e^{\frac{- M^2}{2}}}_{\substack{\propto\  density\ of\\ N(0,1)\ distribution\ !}}
\]
The marginal distribution of $p(\theta,M)$ for $\theta$ is the posterior distribution $\pi(\theta \mid y) \cdot \pi(\theta)$.
So we can sample $(\theta,M)$, and then we only have to look at the values of $\theta$ to get this posterior distribution.
More precisely, to generate proposals for $(\theta,M)$, we proceed as follows:
\begin{itemize}
    \item We draw for our particle a new value of moment $M^*$ from $N(0,1)$.
    \item We then let our particle explore the space $(\theta,M)$ departing from $(\theta,M^*)$, using a \textbf{discretization of the Hamiltonian equations}. After $t$ steps, we reach the state $(\theta^*,M^{**})$, that gives our \textbf{proposal $\theta^*$ for Metropolis-Hastings}.
    \item We then execute the last steps of the Metropolis-Hastings algorithm (Section \ref{MHalgo}) with this proposal.
\end{itemize}

\begin{figure}[H]
    \centering
    \caption{Illustration of HMC inspired from Betancourt (2017)\cite{betancourt2017}}
    \label{Fig::dessinAlgo}
    \includegraphics[height=5cm]{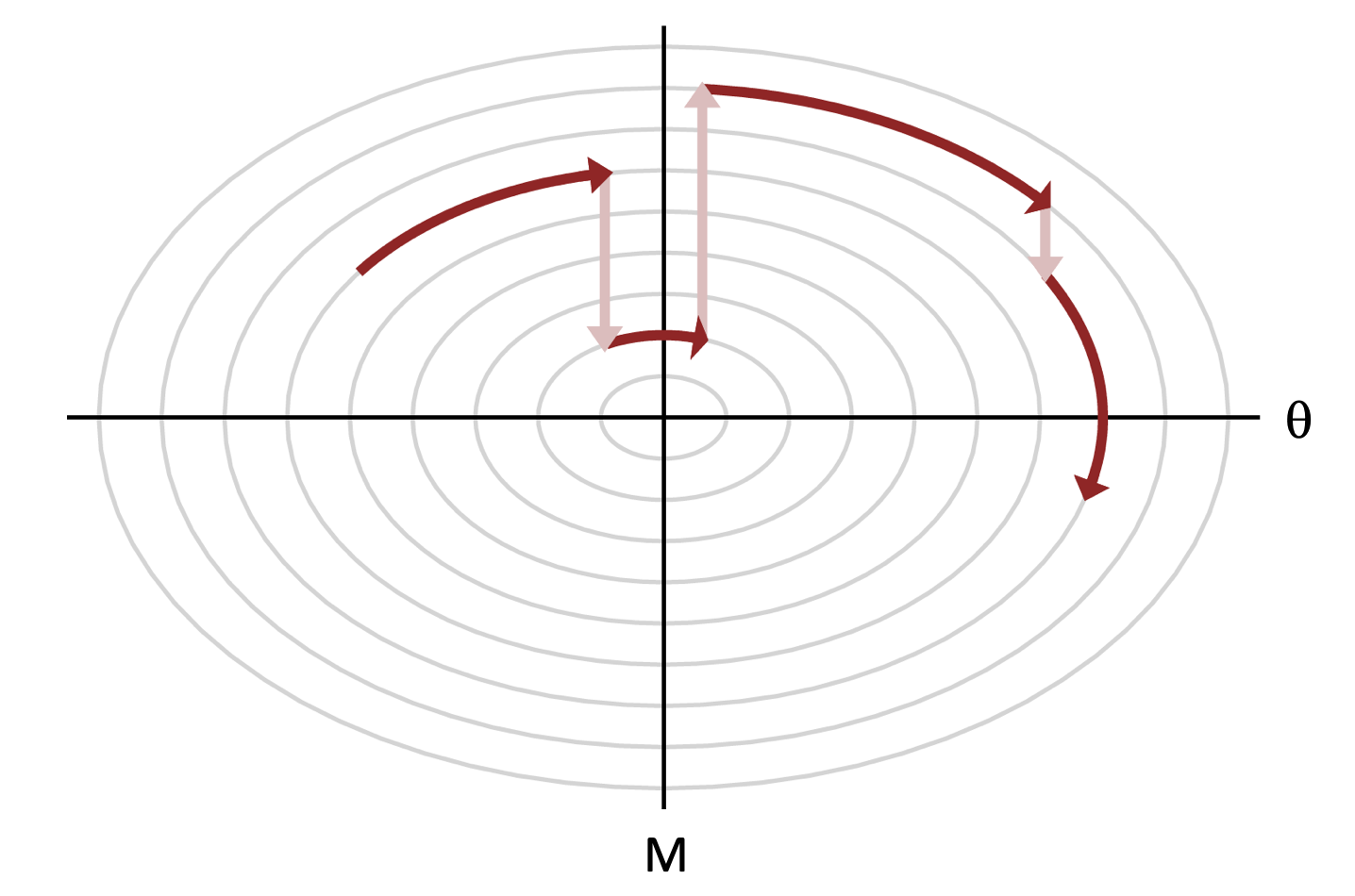}
\end{figure}

Every time we propose a new value of moment $M$, the particle moves up or down to a new energy state, allowing to sample the whole space, as it is illustrated in Figure \ref{Fig::dessinAlgo}.

The aim of this Appendix was to provide the idea behind Hamiltonian Monte Carlo. For further details, I recommend the lecture of the detailed explanations by Lambert (2018)\cite{Lambert}, Betancourt (2017)\cite{betancourt2017} and Neal (2011)\cite{neal2011}.

\end{document}

%% file: setup/packages.tex
\usepackage{titlesec} 
\usepackage[utf8]{inputenc}
\usepackage[x11names,dvipsnames,svgnames,table]{xcolor}

\usepackage[export]{adjustbox}
\usepackage{afterpage}

\usepackage{graphicx}
\usepackage{placeins}
\usepackage{pdfpages}
\usepackage{algorithm2e}
\usepackage{array}
\usepackage{booktabs}
\usepackage[most]{tcolorbox}
\usepackage{calligra}
\usepackage{caption}
\usepackage{datetime}
\usepackage{dblfnote}
\usepackage{dirtytalk}
\usepackage{dsfont}
\usepackage{etex}
\usepackage{fancyhdr}
\usepackage{fix-cm}
\usepackage[T1]{fontenc}
\usepackage{textcomp,gensymb} 
\usepackage{graphicx}
\usepackage{lipsum}
\usepackage{listings}
\usepackage{transparent}
\usepackage[everyline=true,framemethod=tikz]{mdframed}
\usepackage{mparhack}
\usepackage{multicol}
\usepackage{multirow}
\usepackage{parskip}
\usepackage{lscape}
\usepackage{pdflscape}
\usepackage{pdfpages}
\usepackage{placeins}
\usepackage[document]{ragged2e}
\usepackage{rotating}
\usepackage{setspace}
\usepackage{subcaption}
\usepackage{threeparttable}
\usepackage[normalem]{ulem}
\usepackage{verbatim}
\usepackage{soul} 

\usepackage[utf8]{inputenc}
\usepackage[australian]{babel}
\usepackage{csquotes}

\usepackage[a4paper]{geometry}
\usepackage{lastpage} 

\usepackage[hidelinks]{hyperref}
\usepackage{cleveref}

\usepackage{amsmath}
\usepackage{mathtools}

\usepackage{enumitem}

\usepackage[backgroundcolor=yellow]{todonotes}

\usepackage[
    natbib=true,
    style=numeric,
    sorting=none
]{biblatex}

\usepackage[acronym,nonumberlist,toc,section=subsection,numberedsection=nolabel]{glossaries} 
\makeglossaries

\usepackage{float}

\usepackage{amsmath}
\usepackage{relsize}
\usepackage{mathtools}

\usepackage{dsfont}

\usepackage{tocloft}

\usepackage{tcolorbox,listings}
\usepackage{fullpage}
\usepackage{color}
\definecolor{darkWhite}{rgb}{0.94,0.94,0.94}
\lstdefinestyle{frameStyle}{
    basicstyle=\footnotesize,
    numbers=left,
    numbersep=20pt,
    numberstyle=\tiny\color{black}
}
\tcbuselibrary{listings,skins,breakable}
\newtcblisting{customFrame}{
    arc=0mm,
    top=0mm,
    bottom=0mm,
    left=3mm,
    right=0mm,
    width=\textwidth,
    listing only,
    listing options={style=frameStyle},
    breakable
}

\usepackage{amsmath}

\newcommand\myeq{\stackrel{\mathclap{\normalfont\tiny\mbox{def}}}{=}}

\usepackage{titlesec}
\titleformat{\paragraph}
{\normalfont\normalsize\bfseries}{\theparagraph}{1em}{}
\titlespacing*{\paragraph}
{0pt}{3.25ex plus 1ex minus .2ex}{1.5ex plus .2ex}

\usepackage{amsmath}

\usepackage{graphicx}

\usepackage{forest}
\usetikzlibrary{shadows,fit,backgrounds}
\forestset{
  declare toks={level label}{},
  declare toks register={level labels},
  level labels={},
  declare count register={leveller},
  leveller'=0,
  level split/.style={
    temptoksa={#1},
    split register={temptoksa}{:}{tempcounta,level label split},
  },
  level label split/.style={
    temptoksb={#1},
    temptoksc={},
    split register={temptoksb}{,}{temptoksc, level label splitter},
    tikz+/.wrap 2 pgfmath args={
      \node (label leveller ##1) [anchor=east, align=right, font=\sffamily] at (level ##1.west -| westpoint) {##2};
    }{tempcounta}{temptoksc},
    before computing xy/.wrap pgfmath arg={
      tikz+={
        \node [anchor=north east, align=right, font=\sffamily\itshape] at (label leveller ##1.north -| west of westpoint) {Level ##1};
      },
    }{tempcounta},
  },
  level label splitter/.style={
    temptoksc+={\\#1},
  },
  label levels/.style={
    tikz+={
      \coordinate (westpoint) at ([xshift=-15pt]current bounding box.west);
    },
    before packing={
        tikz+={
        \coordinate (west of westpoint) at ([xshift=-15pt]current bounding box.west);
      },
    },
    before drawing tree={
      tikz+={
        \scoped[on background layer]{\node [left color=blue!50!cyan!25!white, right color=blue!50!cyan!25!white, middle color=blue!50!cyan, inner sep=10pt, rounded corners, draw=blue!50!cyan, draw opacity=.5, fill opacity=.15, fit=(current bounding box)] {};}
      },
    },
    delay={
      for tree breadth-first={
        if level label={}{}{
          if={(level())>(leveller)}{
            leveller/.option=level,
            alias/.wrap pgfmath arg={level ##1}{level()},
            if level labels={}{}{
              level labels+={;},
            },
            level labels+/.option=level,
            level labels+={:},
          }{},
          level labels+/.option=level label,
          level labels+={,},
        },
      },
    },
    before typesetting nodes={
      if level labels={}{}{
        split register={level labels}{;}{level split},
      },
    },
  }
}

\usepackage{listings}
\usepackage[most]{tcolorbox}
\usepackage{inconsolata}
\usepackage{graphicx}
\usepackage{xspace,color}
\usepackage{url}
\usepackage{listings}
\usepackage{listings}
\usepackage{xcolor}